\documentclass[fleqn,usenatbib]{mnras}

\usepackage{newtxtext,amsmath}

\usepackage[T1]{fontenc}
\usepackage{ae,aecompl}
\usepackage{color}
\usepackage{tabularx}

\usepackage{graphicx}	
\usepackage{amsmath}	
\usepackage{amssymb}	
\usepackage{overpic}

\title[Short GRBs in the local universe]{Short gamma-ray bursts within 200 Mpc}

\author[Dichiara et al.]{
S. Dichiara$^{1,2}$\thanks{E-mail: \href{mailto:dichiara@umd.edu}{dichiara@umd.edu}},
E. Troja$^{1,2}$,
B. O'Connor$^{1,3,4}$
F. E. Marshall$^{2}$,
P. Beniamini$^{3,4,5}$,\newauthor
J. K. Cannizzo$^{2,6}$,
A. Y. Lien$^{2,6}$
T. Sakamoto$^{7}$ ,
\\
$^{1}$ Department of Astronomy, University of Maryland, College Park, MD 20742-4111, USA \\
$^{2}$ Astrophysics Science 
Division, NASA Goddard Space Flight Center, 8800 Greenbelt Rd, Greenbelt, MD 20771, USA\\
$^{3}$ Department of Physics, The George Washington University, 725 21st Street NW, Washington, DC 20052, USA\\
$^{4}$ Astronomy, Physics and Statistics Institute of Sciences (APSIS)\\
$^{5}$ Division of Physics, Mathematics and Astronomy, California Institute of Technology, Pasadena, CA 91125, USA\\
$^{6}$ Department of Physics, University of Maryland, Baltimore County, 1000 Hilltop Circle, Baltimore, MD 21250, USA \\
$^{7}$ Department of Physics and Mathematics, 
Aoyama Gakuin University, 5-10-1 Fuchinobe, Chuoku, Sagamiharashi Kanagawa 252-5258, Japan 
}
\date{Accepted XXX. Received YYY; in original form ZZZ}

\pubyear{2017}

\begin{document}
\label{firstpage}
\pagerange{\pageref{firstpage}--\pageref{lastpage}}
\maketitle

\begin{abstract}

We present a systematic search for short-duration gamma-ray bursts (GRBs) in the local Universe based on 14 years of observations with the {\it Neil Gehrels  Swift Observatory}. We cross-correlate the GRB positions with the GLADE catalogue of nearby galaxies, and find no event at a distance $\lesssim$100~Mpc and four plausible candidates in the range 100~Mpc$\lesssim$\,$D$\,$\lesssim$200~Mpc. 
Although affected by low statistics, this number is higher than the one expected for chance alignments to random galaxies, and possibly suggests a physical association between these bursts and nearby galaxies. 
By assuming a local origin, we use these events to constrain the range of properties for X-ray counterparts of neutron star mergers. 
Optical upper limits place tight constraints on the 
onset of a blue kilonova, and imply either low masses ($\lesssim10^{-3}\,M_{\odot}$) of lanthanide-poor ejecta or unfavorable orientations ($\theta_{obs}\gtrsim$30~deg).
Finally, we derive that the all-sky rate of detectable short GRBs within 200 Mpc is $1.3^{+1.7}_{-0.8}$ yr$^{-1}$ (68\% confidence interval), and discuss the implications for the GRB outflow structure.
If these candidates are instead of cosmological origin, we set a upper limit of $\lesssim$2.0 yr$^{-1}$ (90\% confidence interval) to the rate of nearby events detectable with operating gamma-ray observatories, such as {\it Swift} and {\it Fermi}. 
\end{abstract}




\section{Introduction}
Short-duration gamma-ray bursts (sGRBs) are sudden and brief flashes of gamma-ray radiation lasting less than 2 seconds \citep{ck93}. Their origin has been traditionally linked to the coalescence of two neutron stars (NSs; \citealt{eichler89,rj99,rosswog03,rezzolla11,jack13}) or a neutron star and a black hole (NS-BH; \citealt{rosswog05,faber06,shibata11}), although no direct proof was found until the historic observations of GW170817 and its gamma-ray counterpart GRB~170817A \citep{LVCGBM, Abbot17m}.  
Consistent with the notion of an old progenitor population, 
\citet{tanvir05} had suggested that a significant fraction (10-25\%) of sGRBs lied in the local Universe, likely harbored in early-type galaxies. 
 These findings, based on crude BATSE localizations of the gamma-ray emission, were not confirmed by the accurate afterglow positions obtained by the {\it Neil Gehrels Swift Observatory} \citep{swift04,gehrels05}.  In 14 years of the {\it Swift} mission, over 100 sGRBs were detected \citep{lien16}, and $\approx$70 were localized to an arcsecond or sub-arcsecond accuracy, yet no event was clearly associated to a nearby ($\lesssim$200 Mpc) galaxy. {\it Swift} observations showed instead that sGRBs reside in all types of galaxy environments spanning a broad range of redshifts, from $z$$\approx$0.1 to $z$$>$2 \citep{berger14}. 
The all-sky rate of sGRBs was consequently revised to a lower value of $\approx$5 Gpc$^{-3}$ yr$^{-1}$ \citep[e.g.][]{cow12,jin15,wp15,ghirlanda2016}.

The discovery of GW170817/GRB170817A at a distance of only 40 Mpc was surprising. It revealed the presence of a local population of faint gamma-ray transients  following NS mergers \citep{LVCGBM}. 
GW170817 was associated with the kilonova AT2017gfo, 
characterized by an early bright optical emission with $M_r\sim$ -16 mag at 12 hours after the merger \citep[e.g.][]{Coulter17}, 
and followed by a delayed afterglow, peaking at $\approx$160~days
after the merger \citep[e.g.][]{troja2017,Hallinan2017,Troja2018b,Mooley18,Resmi18,Margutti18,Lamb19,Davanzo18}.
Another sGRB, GRB 150101B at a distance of $z$=0.1341, was later identified as a GW170817-like explosion with a bright optical kilonova and a late-peaking afterglow \citep{Troja2018a}, showing that this class of transients could be detected by {\it Swift} and ground-based facilities to much larger distances. 
These new results pose the question why local events were not identified before the advent of sensitive GW detectors. 

Observational biases could have played a fundamental role. The typical strategy for the localization of GRB counterparts is based on rapid X-ray observations. However, due to its low-luminosity and delayed onset, an off-axis X-ray afterglow component would be undetectable at early times. Had a GW170817-like event happened during {\it Swift} mission's lifetime, it would probably belong to the sub-sample of {\it Swift} sGRBs with no X-ray counterpart. 
This could explain in part the lack of identifications.

A kilonova component similar to AT2017gfo is instead well above the sensitivity of most ground-based telescopes up to distances of $\approx$200 Mpc. 
Therefore, in the case of a local sGRB, the lack of a kilonova detection appears puzzling. Observations of cosmological sGRBs  place tight constraints on the optical emission for at least some events \citep{gomp18,Rossi2019}, showing that diversity in the kilonova behavior is to be expected. A kilonova fainter or redder than AT2017gfo could have been easily missed in past searches. Such diversity is also expected on theoretical grounds, e. g. a weaker optical emission could characterize
NS-BH mergers or be the signature of NS-NS mergers which promptly collapse to a BH \citep{tanaka14,kawa16}. 
Furthermore, even if an optically bright kilonova was indeed present, the identification of a counterpart embedded within its galaxy light can be a challenging task, especially in the absence of a precise X-ray position.

Based on these considerations, the dearth of nearby sGRBs does not automatically rule out the presence of a local population of faint gamma-ray transients, analogous to GW170817. 
In this paper we address this open question, and explore whether the lack of local sGRBs in the {\it Swift} sample can be mostly ascribed to observing biases or is indicative of a true low rate of nearby events.
A first attempt to identify a group of nearby under-luminous bursts was made by \citet{yue18},
who, however, focused on candidates already reported through GRB Circular Notices and did not perform a comprehensive and homogeneous search over the entire database. 
Systematic searches were recently carried out by \citet{mandhai18} and 
\citet{bartos19} with different sample selection criteria and methodologies. 
\citet{mandhai18} did not find any robust evidence for a population of local sGRBs and constrained their all-sky rate to  $<$4~yr$^{-1}$ within 200 Mpc.  \citet{bartos19} instead proposed a large sample of candidates, 
supporting a rate as high as 10\% of the total sGRB sample \citep{GupteBartos}.
In this work, we provide an alternative strategy for finding local sGRBs 
and present a homogeneous re-analysis of the 
ultraviolet and optical observations in order to constrain the presence of a kilonova, and characterize the contribution of the underlying host galaxy. This is a critical step as the limits and sensitivities typically reported in GRB Circular Notices 
refer to field objects, and do not represent well our ability to detect optical transient sources within bright nearby galaxies. 
The paper is organized as follows: in Section~\ref{sec:data} we present the selected sample of sGRBs, and detail our search strategy for nearby galaxies and data analysis. Results are presented in Section~\ref{sec:results}. In Section~\ref{sec:discussion} we discuss
their implications for the local rate of sGRBs and their outflow structure. 
Conclusions are summarized in Section~\ref{sec:conclusions}. Uncertainties are quoted at the 1$\sigma$ confidence level for each parameter of interest and upper limits are given at a 2 $\sigma$ level, unless stated otherwise. We adopted a standard $\Lambda$CDM cosmology \citep{planck2018}.

\begin{table*}
 	\centering
 	\caption{Short GRBs 
 	with no afterglow detection. 
 	The sample was divided into four sub-groups, depending on the characteristics of the observations (see Section 2.1).  GRBs highlighted in bold-face belong to the ``Gold Sample" of bursts. }
 	\label{tab:sample}
 	\begin{tabular}{lccccccccc}
    \hline
    GRB & $T_{90}$ & Fluence$^{a}$ & 
    R. A. & Dec & 90\% error & SNR & 
    Partial Coding & XRT start & 90\% upper limit \\
         &  [s] & [$\times 10^{-8}$ & (J2000) & (J2000) & [arcmin] &  &  & & [$\times 10^{-14}$ \\ & & erg cm$^{-2}$] & & & & & & & erg cm$^{-2}$ s$^{-1}$] \\  
   \hline
   \multicolumn{10}{c}{Group $a$: Failed to Trigger On-board}  \\
   \hline
    {\bf 080121}$^{b}$ &  $\sim$0.32 & 3$\pm$1 & 09:09:01.8  & +41:50:21.3 & 2.5 & 8.3 & 63\% & 2.3 d & $<$4 \\
    {\bf 090815C} & 0.576$\pm$0.271 & 4$\pm$1 & 04:17:53.5  & -65:54:28.4 & 2.9 & 6.8 & 77\% & 0.4 d & $<$7 \\
    091117 &  $\sim$0.64 &  25$\pm$6  & 02:03:50.0 & -16:56:45.2 & 2.8 & 7.0 & 8\% & 1.1 d & $<$16 \\
    {\bf 100216A}$^{b,c}$ & $\sim$0.208 & 2.6$\pm$0.9 & 10:17:03.1 & +35:31:26.4 & 3.1 & 6.0 & 35\% & 2.5 d & $<$5 \\
    {\bf 100224A} &  0.484$\pm$0.244 & 3.4$\pm$0.8 & 05:33:52.1 & -07:59:38.4 & 2.6 & 7.8 & 100\% & -- & -- \\
   101129A$^{d}$ & 0.350$\pm$0.050 & 13$\pm$4 & 10:23:41.0 & -17:38:42.0 & 3.0 & 6.1 & 12\% & 0.5 d & $<$30 \\     
   120817B$^c$ &  $\sim$0.072 & 17$\pm$4 & 00:33:14.4 & -26:25:40.8 & 2.5 & 8.5 & 5\% & 0.8 d & $<$21 \\   
   {\bf 140402A} & $\sim$0.9 & 5$\pm$2 & 13:50:31.0 & +05:59:41.8 & 2.7 & 7.2 & 66\% & 0.65 d & $<$10 \\ 
   180718A &  0.084$\pm$0.023 & 5$\pm$3 & 22:24:04.6 & +02:47:23.3 & 3.0 & 6.3 & 25\% & 1.3 d & $<$10 \\
    \hline
   \multicolumn{10}{c}{Group $b$: triggered bursts + rapid follow-up}\\
   \hline
{\bf 050906}$^{b,d}$ & 0.128$\pm$0.016 & 1.0$\pm$0.3 & 03:31:13.5 & -14:37:13.1 & 2.8 & 6.9 & 70\% & 79 s & $<$56 \\
{\bf 050925} & 0.092$\pm$0.014 & 7.0$\pm$0.9 & 20:13:56.9 & +34:19:46.1 & 1.5 & 16.8 & 41\% & 72 s & $<$354 \\
{\bf 051105A} & 0.056$\pm$0.014 & 1.6$\pm$0.4 & 17:41:12.2 & +34:56:40.5  & 2.3 & 9.0 & 80\% & 68 s & $<$16 \\
{\bf 070209}$^{b}$ & 0.068$\pm$0.018 & 2.0$\pm$0.4 & 03:04:56.6 & -47:23:16.7 & 2.2 & 9.7 & 85\% & 78 s & $<$16 \\
{\bf 070810B} & 0.072$\pm$0.023 & 1.6$\pm$0.4 & 00:35:49.3 & +08:49:07.8 & 2.3 & 8.5 & 100\% & 62 s & $<$27 \\
{\bf 100628A$^{e}$} & 0.036$\pm$0.008 & 2.0$\pm$0.5 & 15:03:46.3 & -31:39:10.6 & 2.1 & 10.8 & 70\% & 86 s & $<$24 \\
{\bf 130626A} & 0.160$\pm$0.029 & 6.0$\pm$0.8 & 18:12:29.6 & -09:31:29.5   & 1.8 & 13.5 & 100\% & 111 s & $<$1840 \\
{\bf 170112A} & 0.056$\pm$0.018 & 1.3$\pm$0.4 & 01:00:55.7 & -17:13:57.5 & 2.5 & 8.1 & 100\% & 62 s & $<$47 \\
  \hline
   \multicolumn{10}{c}{Group $c$: observing constraints or observed during slew}\\
   \hline
   050202 & 0.112$\pm$0.031 & 3.1$\pm$0.6 & 19:22:20.5 & -38:43:49.4 & 2.2 & 9.7 & 58\%   & -- & -- \\
   070923 & 0.040$\pm$0.009 & 3.9$\pm$0.7 &  12:18:33.9 & -38d:16:52.5 & 1.9 & 12.2 & 77\%   & -- & -- \\
   071112B & 0.304$\pm$0.090 & 4.7$\pm$0.9 & 17:20:47.6 & -80:53:08.5  & 2.3 & 9.4 & 79\% & 62 min & $<$17 \\
   081101 & 0.180$\pm$0.042 & 7.0$\pm$1.1 & 06:23:20.3 & -00:06:17.5  & 1.7 & 13.9 & 51\% & 111 min & $<$4 \\
   090417A & 0.068$\pm$0.021 & 2.2$\pm$0.5 & 02:19:58.4 & -07:08:45.8  & 2.7 & 7.2 & 30\% & -- & -- \\
   110420B & 0.084$\pm$0.021 & 5.5$\pm$1.0 & 21:20:10.9 & -41:16:36.7  & 2.2 & 10 & 36\% & 43 min & $<$15 \\
   111126A &  0.672$\pm$0.039 & 7$\pm$1 & 18:24:07.1 & +51:28:06.1 & 2.5 & 8.3 & slew & -- & -- \\
   120229A & 0.236$\pm$0.037 & 4.1$\pm$0.7 &  01:20:09.8 & -35:47:54.4  & 1.9 & 12.5 & 86\% & -- & -- \\
   140414A & 0.496$\pm$0.068 & 12.4$\pm$1.2 &  13:01:20.3 & +56d:54:41.8  & 1.7 & 13.7 & slew & 11.7 hr & $<$60 \\
   140606A & 0.340$\pm$0.094 & 5.1$\pm$1.0 & 13:27:11.7 & +37:35:55.8  & 2.4 & 8.6 & 87\% & -- & -- \\
   151228A & 0.276$\pm$0.040 & 8.3$\pm$1.1 &   14:16:04.8 & -17:39:59.1   & 1.8 & 13.3 & 100\% & 2.1 d & $<$24 \\
   160612A & 0.248$\pm$0.048 & 9.6$\pm$1.0 & 23:13:27.3 &  -25:22:28.1 & 1.7 & 13.9  & slew & 2.1 d & $<$31 \\
   160726A & 0.728$\pm$0.043 & 27.2$\pm$2.4 & 06:35:14.3 & -06:37:01.4  & 1.3 & 21.2 & 56\% & -- & -- \\
   170325A & 0.332$\pm$0.071 & 8.6$\pm$1.3 &   08:29:55.9 & +20:31:32.5  & 2.0 & 11.4 & slew & -- & -- \\
   180715A & 0.684$\pm$0.088 & 12.6$\pm$1.9 & 15:40:20.5 & -00:53:57.5  & 2.0 & 11.3 & 85\% & 53 min & $<$31\\
     \hline
      \multicolumn{10}{c}{Group $d$: INTEGRAL bursts}\\
   \hline
   081226B & $\sim$0.5 & $\sim$10 & 01:41:58.8 & -47:26:20 & 2.2 & -- & --  & 2.7 hr & $<$21 \\
   110112B & $\sim$0.3 & $\sim$10 & 00:42:23.8 & +64:24:22.6  & 2.6 & -- & -- & 15.5 hr & $<$30 \\
   131224A & $\sim$0.8 & $\sim$3 & 19:47:20 & +31:40:03 & 2.0 & -- & -- & 3 hr & $<$28 \\
    \hline
    \hline
    \end{tabular}
\begin{flushleft}
\quad \footnotesize{$^a$ Fluence calculated in the 15-150 keV energy range for {\it Swift} GRBs \citep{lien16} and 20-200 keV for INTEGRAL GRBs.}\\
\quad \footnotesize{$^b$ Bursts with a GLADE catalog local galaxy (D$\lesssim$200 Mpc) inside the 90\% error region}\\
\quad \footnotesize{$^c$ Standard tools fail to estimate the $T_{90}$ error for these bursts. 
}\\
\quad \footnotesize{$^d$ Standard tools fail to estimate the $T_{90}$ value. $T_{90}$ is retrieved from GCNs \citep{gcn3935,gcn11436}.}\\ 
\quad \footnotesize{$^e$  The X-ray candidate counterpart shows only 7 total counts with 0.7 of expected background. This yields a low statistical significance, below what we define as detection threshold.}
\end{flushleft}
\end{table*}

\section{Data Analysis}\label{sec:data}

\subsection{Sample selection}

\subsubsection{{\it Swift} bursts}
We considered all the events detected by {\it Swift} between January 1st, 2005 and January 1st, 2019, as reported in the online BAT GRB catalogue\footnote{https://swift.gsfc.nasa.gov/results/batgrbcat/} \citep{lien16}.
We found 119 events classified as sGRBs ($T_{90}$\,$<$\,2~s),
with 37 events not associated with any X-ray or optical counterparts. 
We did not include sGRBs with extended emission \citep{Norris06}, whose classification and progenitor are still highly uncertain.
In addition, since the traditional 2-s cut might introduce a significant contamination from the population of long GRBs \citep{Bromberg13}, 
we adopted a stricter selection criterion of $T_{90}$\,$<$\,1~s. This choice reduces the number of interlopers without significantly impacting the sample size. 
Out of 96 sGRBs with $T_{90}$\,$<$\,1~s,
we select 32 events without afterglow counterpart\footnote{
GRB070406, although reported in the BAT catalogue, was removed from the analysis due to the low significance ($\sim$4~$\sigma$) of the signal in BAT.}.
Table~\ref{tab:sample} lists their properties. 
Prompt emission properties were derived from the {\it Swift}/BAT data using standard analysis procedures \citep{lien16}. {\it Swift}/XRT upper limits
were derived using the online {\it Swift} tool\footnote{http://www.swift.ac.uk/user\_objects/}, and converted into fluxes using a conversion factor of 4$\times$10$^{-11}$ erg\,cm$^{-2}$\, cts$^{-1}$, typical of GRB afterglows \citep{evans09}.
This sample can be divided into three sub-groups:\\
\indent {\it Group a}: these events (9 in total) were not detected through the on-board trigger algorithm, and were found at a later time by a human-based analysis. 
As can be derived from Table~\ref{tab:sample}, 
the lack of on-board detection is either due to a low $\gamma$-ray fluence in the BAT bandpass (intrinsic factor) or to an unfavorable position (extrinsic factor), at the edges of the BAT field of view (partial coding $\lesssim$25\%). \\
\indent{\it Group b}: these events (8 in total) were triggered normally and rapidly ($\lesssim$1000~s) observed with {\it Swift}'s narrow field instruments, yet no afterglow was found. \\
\indent {\it Group c}: about half of the bursts with no afterglow counterpart (15 in total) were affected by extrinsic factors,
such as observing constraints or a spacecraft's slew, which delayed their identification and observations. Therefore this group include bursts triggered on-board but without early afterglow observations and bursts which were not triggered the on-board because observed during a slew.

\subsubsection{Other missions}

Short GRBs detected by the {\it Fermi} Gamma-ray Telescope or through the InterPlanetary Network (IPN) are characterized by large error regions, and are not well suited for our study. 
We include in our selection bursts localized by the INTEGRAL Soft Gamma-Ray Imager\footnote{https://www.isdc.unige.ch/integral/science/grb\#ISGRI} (ISGRI; \citealt{Lebrun03}), with a typical localization uncertainty of $\lesssim$3 arcmin. 
We found 5 events classified as short bursts:
GRB 150831A, also triggered by {\it Swift}, and GRB 070707 
have an afterglow counterpart, 
whereas the three remaining bursts (GRB 131224A, GRB 110112B and GRB 081226B) have no detected afterglow. 
This fourth group (\textit{Group d}) is also listed in Table~\ref{tab:sample}. 

The four sub-groups identified in Table~1 form our ``Total Sample" of bursts.
External factors (e.g. disabled trigger during the slew, proximity to the Sun) often affected their observations. 
In these cases, bursts were likely drawn from the standard population of cosmological sGRBs, and the lack of an afterglow counterpart could simply be the result of sub-optimal observations.
After excluding these events, we define a ``Gold Sample'' of 13 bursts (bold-face in Table 1) observed in optimal conditions, yet not detected by {\it Swift}. 
We consider that this sample is more likely to intrinsically differ from the general population.
These are bursts which had a favorable sky position (partial coding $>$25\%) and failed to trigger BAT (group $a$), or bursts
whose afterglow was not detected despite rapid ($\lesssim$1,000~s) follow-up observations (group $b$).

\subsection{Search for nearby galaxies}

We cross-correlated the burst positions in Table~1 with the GLADE v2.3 catalogue of nearby galaxies \citep{glade}. 
GLADE contains $\approx$3 million objects classified as galaxies and is complete in terms of their measured B-band luminosity up to $\approx$40~Mpc. Its completeness decreases to $\approx$40\% at 200 Mpc.  
SGRBs typically reside in bright galaxies \citep{Gehrels2016} and, by considering only the brighter half of galaxies,  the completeness of the catalogue increases to $\gtrsim$90\% up to 200 Mpc. 

We considered a match if we find one or more nearby galaxies (estimated distance of $\lesssim$200~Mpc) within the quoted 90\% error region. 
Four matches were found out of the 13 bursts in our ``Gold Sample'', corresponding to 31\%. These matches are GRB 050906, GRB 080121, GRB 070810B and GRB 100216A
No other match is found by including the remaining 22 bursts of the ``Total Sample''. We note that a bright ($B \sim$17.01 mag) galaxy HyperLEDA PGC890767 lies within the error region of the BAT GRB~091117 and
a galaxy
HyperLEDA PGC501449 ($B \sim$18.6 mag) lies within the error region of the INTEGRAL GRB~081226B. 
However, no information on their distances are available, and we do not include them in our selection.

We repeated the same analysis using the BAT positions from the sample of sGRBs with an XRT localization, and from the sample of long GRBs. In both cases, we found a significantly lower match rate\footnote{For completeness
we report that two sGRBs with X-ray afterglow match with a nearby GLADE galaxy: GRB~150101A and GRB~070809, as noted by \citet{Tunnicliffe2014}.
However, the optical properties of GRB~070809  seem to favor $z\gtrsim$0.2 \citep{Jin2019}.}
($\sim$3-4\%).
Since the accuracy of the BAT localization depends on the signal-to-noise ratio, weak bursts have less accurate positions than average. We tested whether the higher match rate of the ``Gold Sample'' of bursts was a consequence of their lower fluences and hence larger error regions. We simulated 1,000 random sky positions with the same localization accuracy of our ``Gold Sample''
and found that the match rate remains low ($\sim$3\%), confirming that the size of the error region does not drive the result. 

The  match rate between sGRBs with no X-ray afterglow and nearby galaxies is therefore higher than the rate expected from chance alignments, possibly as a result of a real physical connection between some of these bursts and nearby galaxies. Due to the low number of events, this excess is only marginally significant ($\approx$98\%) at a statistical level \citep{Gehrels1986}.

\subsection{Optical data reduction}

We examined in detail the follow-up observations for the four sGRBs possibly associated to nearby galaxies. 
Our dataset includes archival data from the
UltraViolet and Optical Telescope \citep[UVOT; ][]{roming2005} aboard  {\it Swift},
the Low Resolution Imaging Spectrometer (LRIS; \citealt{oke1995}) 
on the 10~m Keck I telescope, 
and the Gemini Multi-Object Spectrographs (GMOS; \citealt{hook2004})
on the 8.1~m Gemini-North telescope. 

{\it Swift} data were reduced in a standard fashion using the HEASOFT package v6.26.1 and the latest calibration files. Aperture photometry was performed on the images using the prescriptions of \citet{Breeveld2010}.

Ground-based imaging data were reduced using standard techniques for CCD data reduction, including e.g. bias subtraction, cosmic-ray rejection, flat-field correction. 
We used custom IDL scripts and the standard Gemini 
IRAF\footnote{IRAF is distributed by the National Optical Astronomy Observatory, which is operated by the Association of Universities for Research in Astronomy (AURA) under a cooperative agreement with the National Science Foundation.} package. 
If the target was observed for two or more epochs, we searched for variability by performing image subtraction with HOTPANTS \citep{Becker2015}.
Magnitudes were calculated using aperture photometry and calibrated against the SDSS DR15 catalogue \citep{Blanton2017}.
Upper limits were derived by seeding the images with artificial point-like sources of known brightness.

\begin{figure}
\includegraphics[scale=0.72,trim=2 3 27 13, clip]{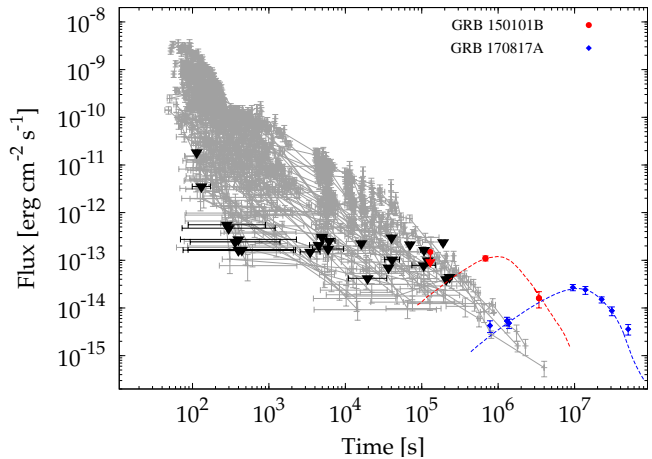}
\caption{Comparison between the sample of {\it Swift} sGRBs with X-ray afterglows and our upper limits. Red circles and blue diamonds show the X-ray afterglows for GRB 150101B \citep{Troja2018a} and GRB 170817A \citep{Troja2019a}, respectively. Dashed lines show the off-axis afterglow models. 
\label{xray_limits_sGRBs}}
\end{figure}

\section{Results}\label{sec:results}
\subsection{X-ray constraints}

XRT upper limits on the early afterglow phase were derived using the online {\it Swift} tool\footnote{http://www.swift.ac.uk/user\_objects/},
and are reported in Table~\ref{tab:sample}. 

In order to estimate the typical sensitivity of the {\it Swift} observations we derive the upper limit at a single position within the BAT error region referred to the first snapshot taken by the instrument after the detection.
They may differ from the values reported in GRB Circular Notices, often derived using longer integration times.
 
In Figure~\ref{xray_limits_sGRBs}  we compare these X-ray limits (3-$\sigma$ c.l.) 
to the {\it Swift} sample of sGRB afterglows, and to the off-axis X-ray afterglows of GW170817/GRB170817A and GRB150101B detected by {\it Chandra} \citep{Troja2018a,Troja2019a}.
For early follow-up ($\lesssim$6 hrs) observations with no XRT detection, the X-ray afterglow must be fainter than the average sGRB population, although still within the distribution of observed X-ray fluxes. Upper limits become less constraining in the case of delayed ($\gtrsim$12 hrs) observations. In no case were {\it Swift} observations sensitive to late-peaking afterglows, such as GW170817. 

\begin{figure}
\includegraphics[scale=0.765,trim=2 3 27 13]{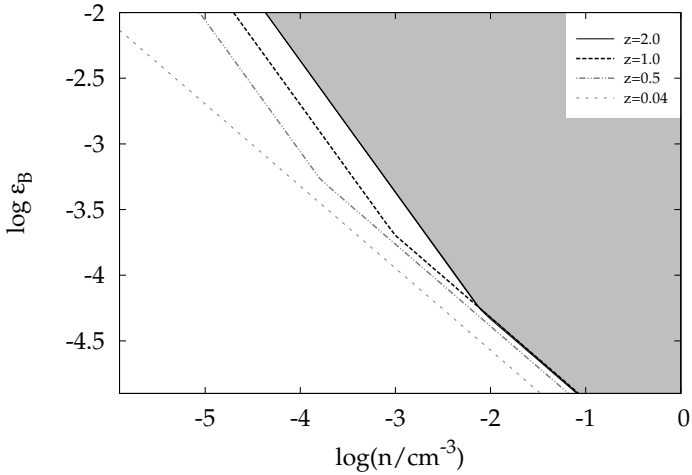}
\caption{The allowed parameters for non-detection in X-rays for an upper limit of $F_{X,\textrm{lim}}=1.6\times 10^{-13}\,\textrm{erg }\textrm{cm}^{-2}\textrm{s}^{-1}$ at early times (i.e. $70-2200$ s) for a GRB with the average kinetic energy (scaled to each redshift) of the four bursts with hosts identified at $d<200$ Mpc within this paper. The allowed parameter space for each redshift is to the left of the line defining the region. The shaded gray region represents the disallowed parameter space for $z\leq 2$. 
\label{xray_allowed_regions}}
\end{figure}

The early XRT upper limits can be used to constrain the properties of an on-axis GRB explosion. In order to do this, we explore the allowed parameter space that would lead to a non-detection at early times. 
We apply the standard model of synchrotron afterglow emission \citep{Meszaros1997,Sari1998} using the formulation of \citet{Granot2002} for a constant density interstellar medium (ISM), and including Inverse Compton (IC) corrections \citep{Sari2001,Zou2009,Beniamini2015}. 
This model is completely described by a set of five parameters: \{$p$, $\varepsilon_B$, $\varepsilon_e$, $E_\textrm{kin}$, $n$ \}, where $p$ is the slope of the electrons' power-law energy distribution, $\varepsilon_B$ and $\varepsilon_e$ are the fractions of the burst kinetic energy $E_\textrm{kin}$ that exist in the magnetic field and electrons respectively, and $n$ is the circumburst particle density. 
We adopt $p=2.2$ and $\varepsilon_e=0.1$ as canonical values \citep{BeniaminiVanDerHorst}. 
We also assumed a gamma-ray efficiency of $\eta_\gamma=0.2$ \citep{Beniamini2015} to convert the gamma-ray energy released
into blastwave kinetic energy $E_\textrm{kin}$,
and an outflow Lorentz factor $\Gamma$=300 \citep{Ghirlanda2017}.

The allowed parameter space is defined by requiring that
the X-ray flux is below the XRT sensitivity ($\approx$2$\times10^{-13}$ erg cm$^{-2}$ s$^{-1}$) at the time of first observation ($\approx$70~s). This yields two conditions:
\begin{align}
    \varepsilon_B\, n^{5/8} < 4.6\times 10^{-8}\, d_{L,28}^{5/2} E_{\gamma,\textrm{iso},52}^{-13/8} \eta_{\gamma}^{13/8} (1+z)^{-13/8},
\end{align}
\begin{align}
    \varepsilon_B\, n < 3.3\times 10^{-9}\, d_{L,28}^{5/2}E_{\gamma,\textrm{iso},52}^{-5/4} \eta_{\gamma}^{5/4} (1+z)^{-1/2},
\end{align}
where $d_{L,28}$ is the luminosity distance.
The true parameter space for non-detections is the maximum allowed value of $\varepsilon_B$ and $n$ from the two regions, as shown in Figure \ref{xray_allowed_regions}. 
We find that, at any redshift, the scenario of on-axis sGRBs is consistent with the lack of X-ray detection if these explosions happened in a tenuous environment, $n \lesssim 4 \times 10^{-3}$\,cm$^{-3}$ for $\epsilon_B \gtrsim 10^{-4}$. The density can be greater than  $n\approx10^{-1}$ cm$^{-3}$ only for $\epsilon_B\lesssim 10^{-5}$. 

\begin{figure}
\includegraphics[scale=0.72, trim=2 3 27 13,]{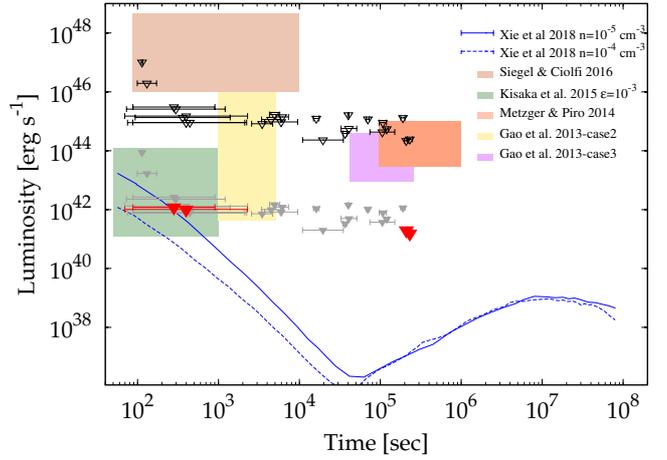}
\caption{XRT upper limits for the selected sample of bursts compared with different models for X-ray counterparts of NS mergers. Filled and empty symbols are the upper limits at 200 Mpc and $z$=1, respectively. 
Red downward triangles are the limits derived for GRB050906, GRB080121, GRB070810B, and GRB100216A using the distance of their candidate host galaxy. 
\label{xray_limits}}
\end{figure}

We also compare the X-ray upper limits with models of alternative X-ray counterparts to NS mergers (Figure~\ref{xray_limits}). 
A first set of models discuss the emission from a long-lived and highly magnetized NS remnant. \citet{Gao2013}, \citet{Siegel2016} and \citet{Metzger2014}
predict bright X-ray counterparts, which can be ruled out by the XRT upper limits assuming a local origin ($d_L<$200Mpc) for the short GRBs in our sample.
The predictions of \cite{Siegel2016} fall above our limits even assuming $z$=1 for all the busts (open downward triangles in Figure~\ref{xray_limits}).

A second set of models consider the X-ray emission arising from the interaction between the relativistic jet and the merger ejecta \citep{Kisaka2015,Xie2018}, and predict faint signals that peak at early times. These are only weakly constrained by the XRT limits.
\cite{Kisaka2015} propose that photons emitted from the jet are scattered at large angles by the surrounding ejecta, producing a nearly isotropic X-ray transient. XRT limits can exclude part of the parameter space, but remain consistent with this model for low on-axis X-ray luminosity ($L_{X, iso}\lesssim10^{45}$ erg s$^{-1}$) or a low scattering parameter ($\epsilon\lesssim10^{-5}$).
\cite{Xie2018} discuss instead the possibility of an X-ray flash immediately after the merger. When the relativistic jet emerges from the cloud of ejecta, a shock-heated layer of mildly relativistic material ($\Gamma<$2) radiates non-thermally in the X-ray band. This emission is below our upper limits for typical densities of the external medium ($n>10^{-5}$\,cm$^{-3}$).

\begin{center}
\begin{figure*}
\includegraphics[scale=0.74]{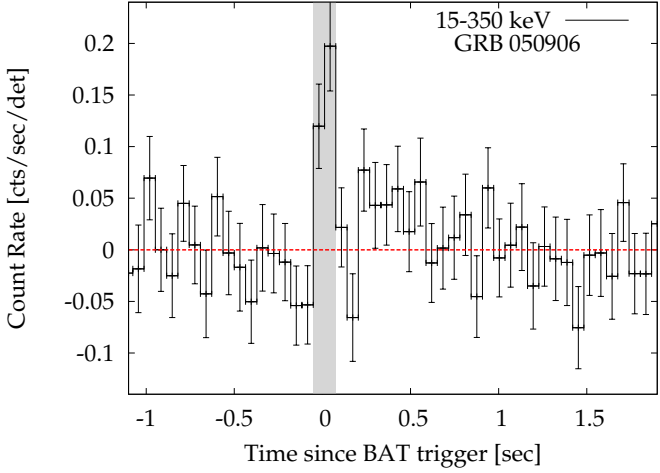}
\includegraphics[scale=0.42, bb=110 50 660 490, clip=true]{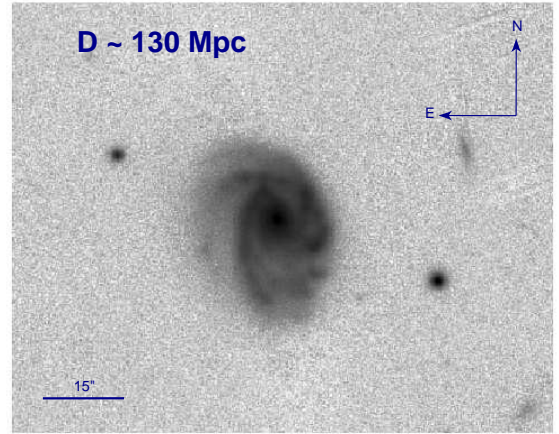}
\caption{Left:\textit{Swift}/BAT mask-weighted light curves of GRB 050906 in the energy range 15-350 keV. The time bin is 64 ms. The vertical bar shows the $T_{90}$ interval. Right: PannSTARRS r band image of the candidate host galaxy IC328.     
\label{batlc-1}}
\end{figure*}
\end{center}

\begin{center}
\begin{figure*}
\includegraphics[scale=0.74]{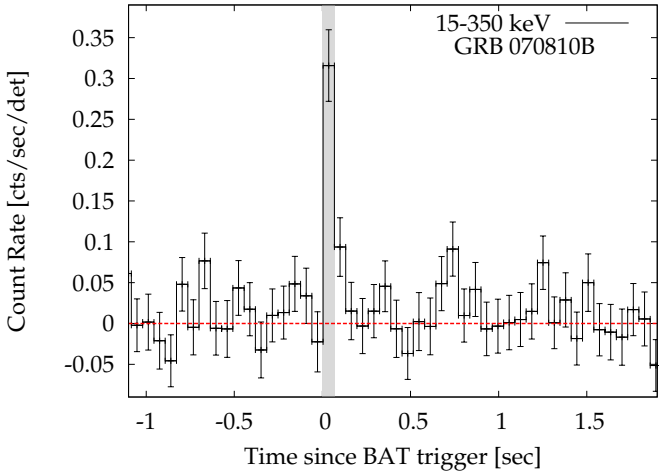}
\includegraphics[scale=0.42, bb=110 50 660 490, clip=true]{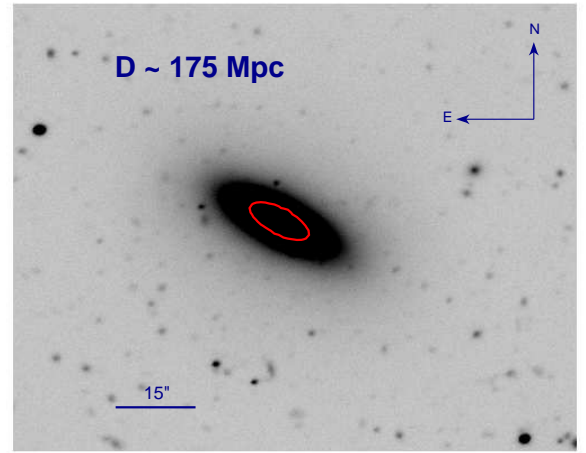}
\caption{Left: GRB 070810B BAT mask-weighted light curve (15-350 keV), with 64 ms time bin. The vertical box shows the $T_{90}$ interval. Right: Keck r-band image of the candidate host galaxy 2MASX J00355339+0849273
at $\sim$175~Mpc.
The red contour shows the inner region where no optical source can be reliably detected (saturated region). The probability to find a sGRB at a similar offset from the galaxy's center is $\lesssim$18\%  \citep{berger14}. 
\label{batlc-2}}
\end{figure*}
\end{center}

\begin{center}
\begin{figure*}
\includegraphics[scale=0.74]{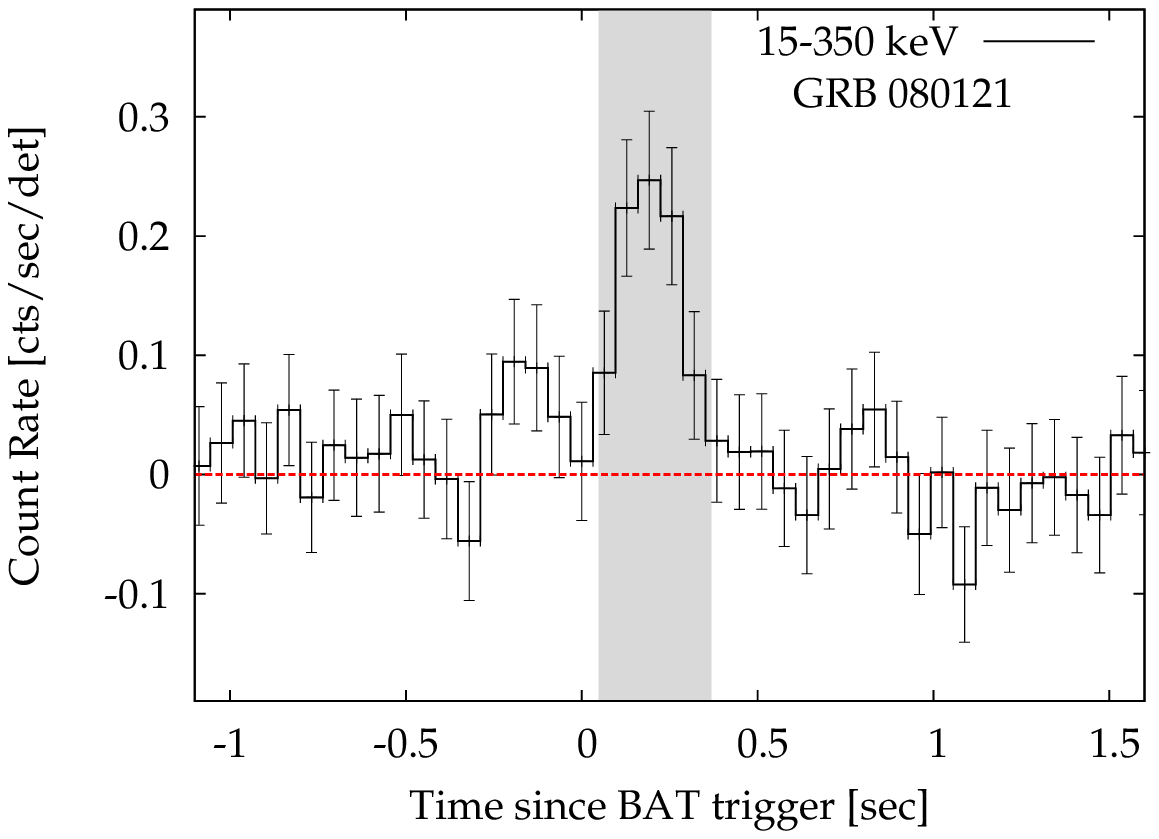}\llap{\makebox[0pt][l]{\hspace*{-7.8cm}\raisebox{+2.0\height}{\includegraphics[scale=0.237]{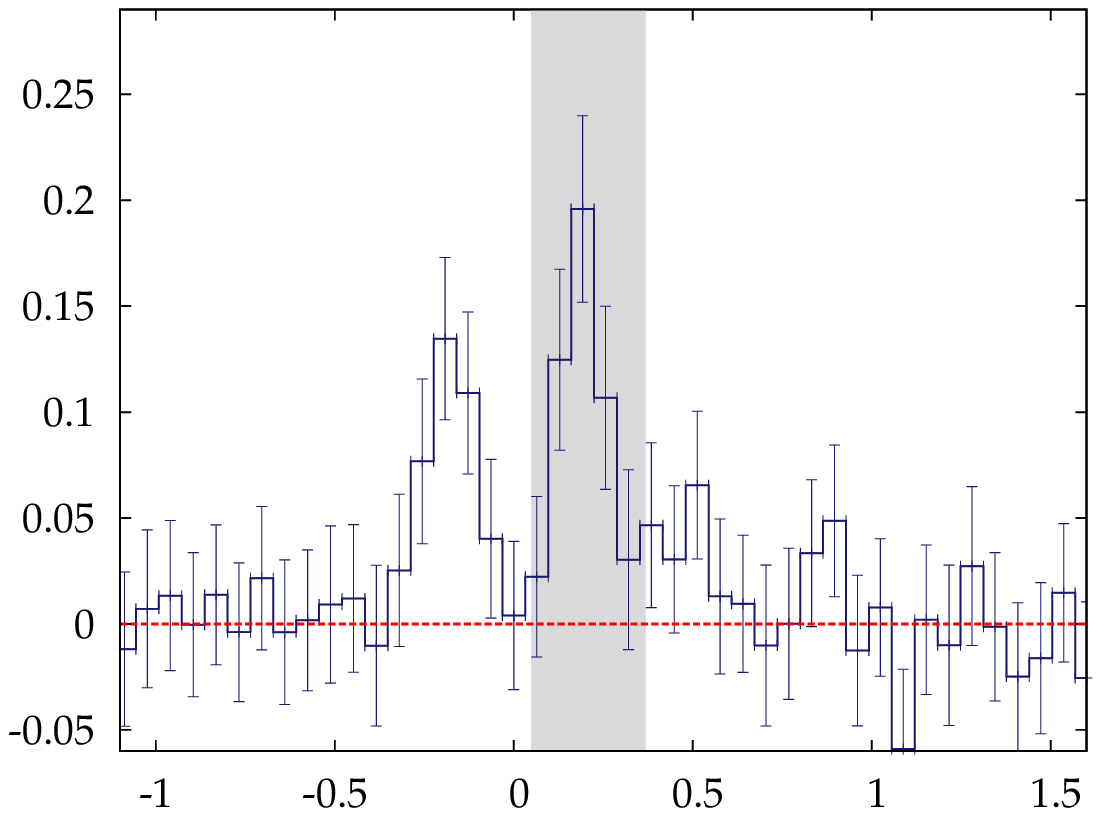}}}}
\includegraphics[scale=0.42, bb=110 50 660 490, clip=true]{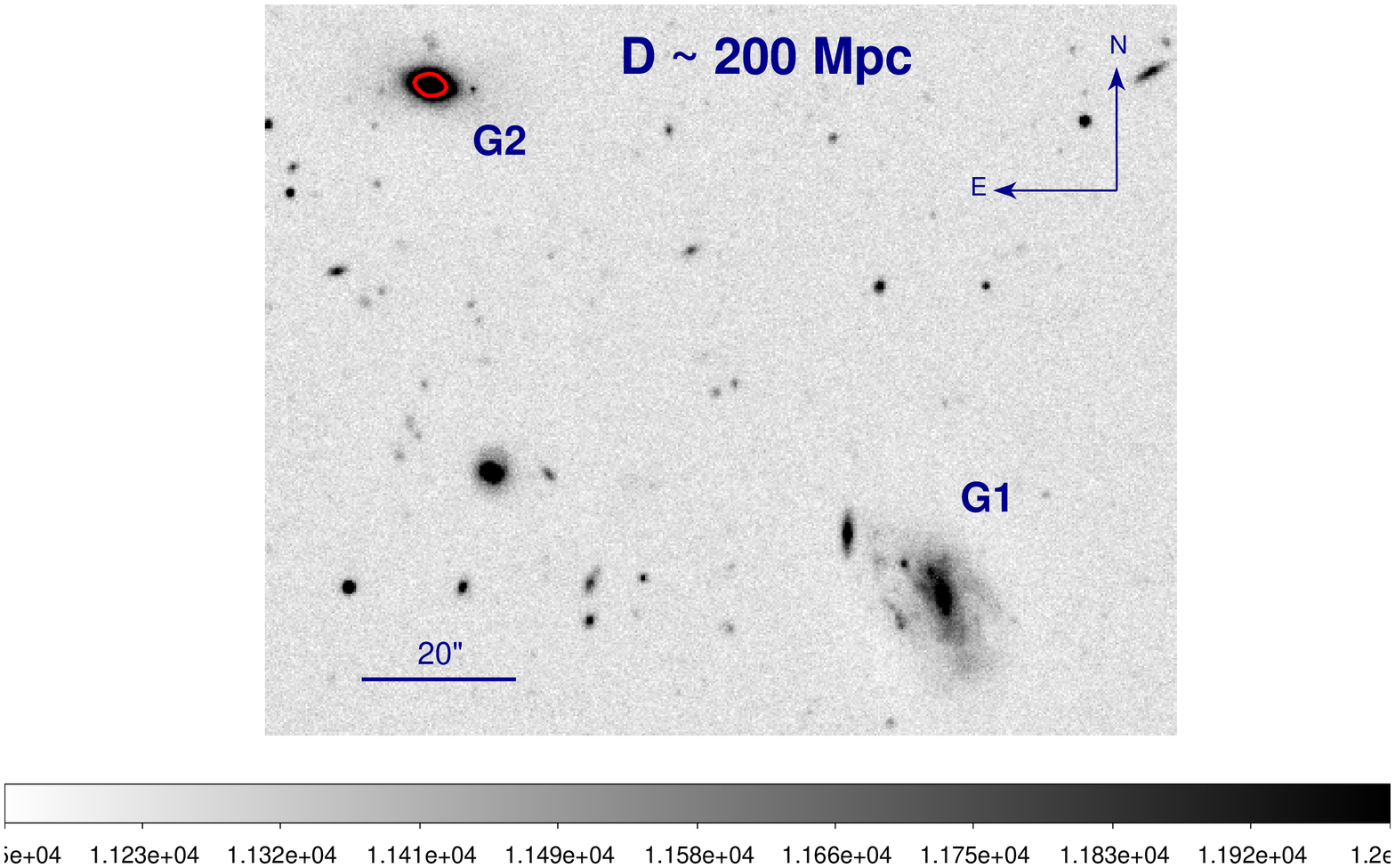}
\caption{Left: 
GRB 080121 light curve with 64 ms binning scheme (15-350 keV). 
The inset shows the temporal profile at lower energies (15-50 keV),
where a precursor is visible. The $T_{90}$ interval is shown by the vertical bar. 
Right: The two possible host galaxies are shown in the Gemini r filter image. G1 and G2 are SDSS J090858.15+414926.5 and SDSS J090904.12+415033.2, respectively. 
The overlaid contour on G2 shows the inner region where no optical source can be reliably detected (saturated region). The probability for a sGRB to occur inside this region is $\lesssim$13\% \citep{berger14}.
\label{batlc-3}}
\end{figure*}
\end{center}

\begin{center}
\begin{figure*}
\includegraphics[scale=0.74]{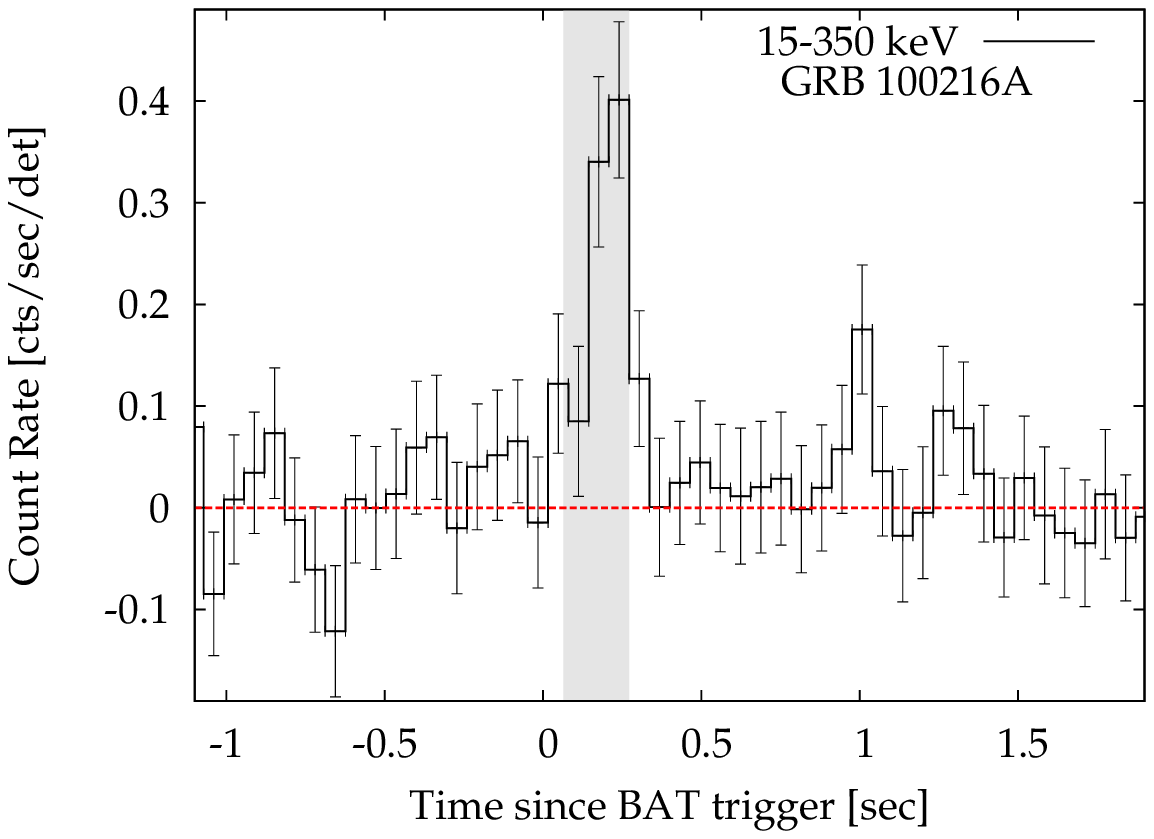}
\includegraphics[scale=0.42, bb=110 50 660 490, clip=true]{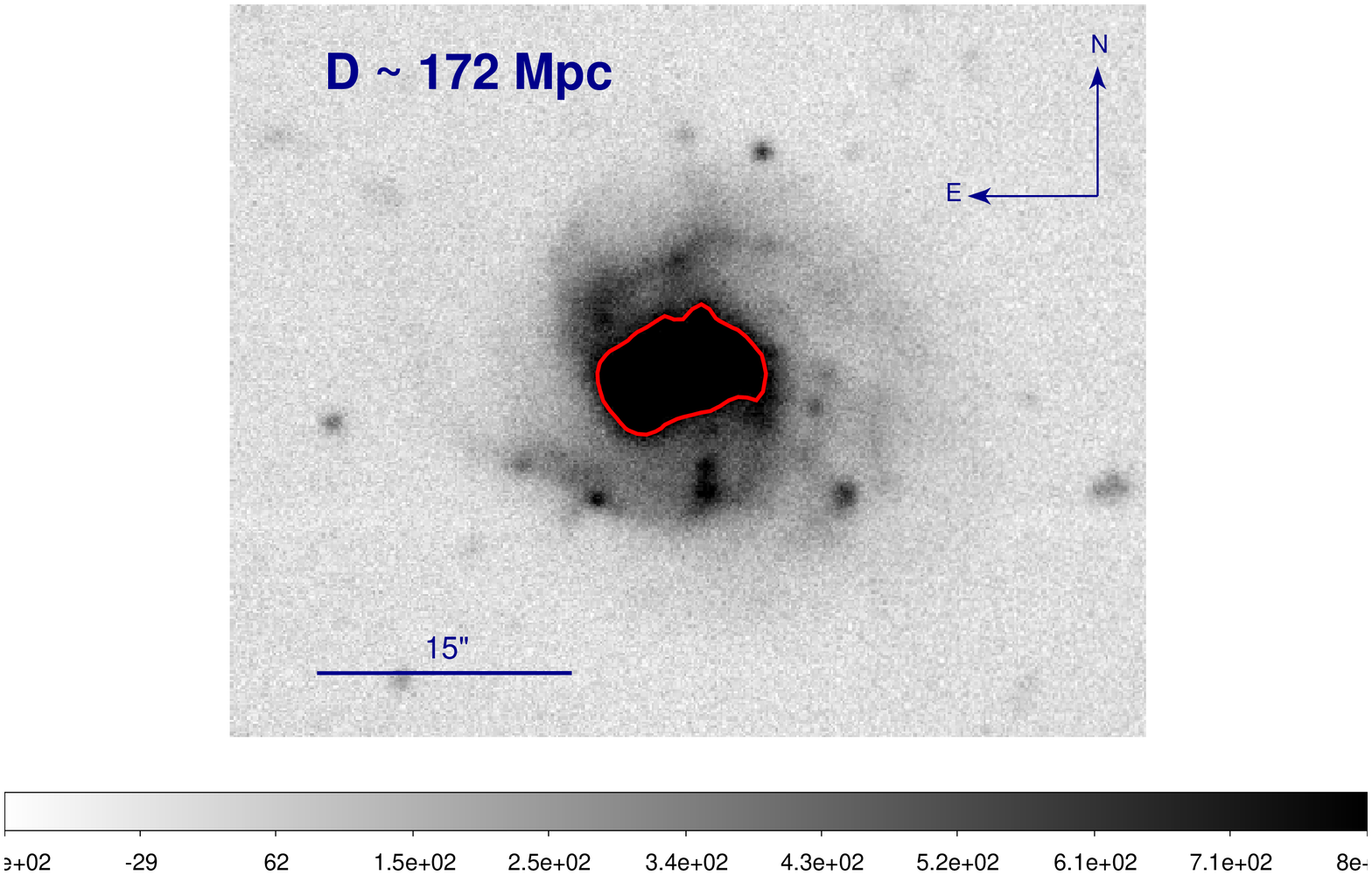}
\caption{Left: GRB 100216A BAT light curve with 64 ms binning scheme (15-350 keV). The $T_{90}$ interval is shown by the vertical bar. Right: Keck r-band image of the putative host galaxy. 
The overlaid contour shows the inner region where no optical transient source could be identified. We could not perform image subtraction in this case, since the frames were collected during a single observational epoch. We used PanSTARRS reference image to investigate the presence of possible transients. The probability for a sGRB to occur inside this region is $\lesssim$35\% \citep{berger14}.}
\label{batlc-4}
\end{figure*}
\end{center}

\subsection{Candidate events within 200 Mpc}
The four matches found in the ``Gold sample'' are GRB050906, GRB080121, GRB070810B, and GRB100216A. Their prompt emission is shown in Figure~\ref{batlc-1}--\ref{batlc-4} 
together with their putative host galaxies. 
Below we discuss in detail their properties and compare them with the electromagnetic counterparts of GW170817.

\subsubsection{GRB050906}
This burst was studied in detail by \cite{levan08} who noted the proximity with IC328 and the companion IC327 at redshift $z$=0.031 ($\sim$134~Mpc). 
Whereas the two galaxies lied outside the BAT position used in \cite{levan08}, IC328 is fully consistent with the updated position reported in the 3rd BAT catalogue \citep{lien16}. The image of the galaxy within the updated BAT error region is available in the online supplementary material. The maximum distance between the center of the galaxy and the border of the 90\% BAT error region is $\sim$240$\arcsec$ which corresponds to a maximum offset of $\sim$150 kpc.

The GRB prompt emission is marginally detected by BAT
(it was at the limit of the trigger threshold with a SNR=6.9, Figure~\ref{batlc-1}) and not particularly hard in spectrum. 
At a distance of $\approx$130 Mpc, 
its energy release would be $\approx$2$\times$10$^{46}$ erg.
Despite the close distance and the rapid follow-up, no afterglow counterpart was found. 
\cite{levan08} suggested that this unusual properties could be
evidence of an extra-galactic soft gamma-ray repeater. Instead, here
we discuss whether they could be consistent with a NS merger. 

If IC328 is indeed the host galaxy of GRB~050906, then the burst environment is markedly different from GW170817. The galaxy morphology
indicates a late-type galaxy, its colors suggest a relatively high stellar mass, $M \approx 10^{11} M_{\odot}$, and a moderate star formation rate log(SFR/M$_{\odot}$ yr$^{-1})$ $\approx$ 1.2 \citep{levan08}.
Although different from NGC~4993 \citep[e.g.][]{Im17}, 
these global properties are consistent with the heterogeneous environment of sGRBs \citep{berger14}.

No lower limit for compact object mergers is available for this GRB since LIGO was not operative at the time of the burst.

\subsubsection{GRB070810B}

The prompt phase observed by BAT consist of a very short ($T_{90}$=0.07~s) single pulse (Figure~\ref{batlc-2}).
XRT observations started 62~s after the trigger, but no reliable X-ray afterglow was identified. No optical counterpart is found in the early UVOT observations, starting 65 seconds after the trigger \citep{gcn070810b}. 

A bright early-type galaxy (Figure~\ref{batlc-2}) at $z$=0.0385 \citep[$\sim$175 Mpc; ][]{thoene07} lies within the BAT error circle. The maximum projected offset, derived considering the border of the 90\% BAT error region, would be  $\sim$215$\arcsec$, corresponding to $\sim$175 kpc

At this distance, the burst isotropic-equivalent energy would be $E_{\gamma,iso} \approx 6\times10^{46}$~erg in 15--150 keV.
The putative host is an evolved spheroidal galaxy, detected by 
the Wide-field Infrared Survey Explorer (WISE;
\citealt{cutri2013wise}) with W1$(3.4 \mu m)$=14.79$\pm$0.02 AB mag, and W2$(4.6 \mu m)$=15.47$\pm$0.03 AB mag .
Using the relation of \cite{wen2013}, we infer a stellar mass of $log(M/M_{\odot})\approx$ 9.8. 
From UV observations with the GALaxy Evolution EXplorer (GALEX; \citealt{Bianchi17}), 
this galaxy has a magnitude of 23.01$\pm$0.17 AB mag and 20.93$\pm$0.06 AB mag in the far and near UV band, respectively. Using the relations provided by \cite{iglesias2006}, we derive a star formation rate of log(SFR/M$_{\odot}$ yr$^{-1})$ $\sim$ 0.21 (see the complete table included in the online supplementary material). 
The galaxy's morphology, its large stellar mass and low star formation are
similar to the environment of GW170817, and other nearby sGRBs such as GRB~150101B \citep{Troja2018a} and GRB050724A \citep{berger14}. 

This burst happened during LIGO's fifth science run (S5). Analysis of the GW data could only exclude that the merger happened within 2 Mpc for a NS-NS merger, and 6 Mpc for a NS-BH merger (90\% confidence level; \cite{Abadie2010}).

 \subsubsection{GRB080121}

GRB 080121 was discovered through the ground-based analysis of {\it Swift}/BAT data \citep{gcn7209}. Its prompt gamma-ray phase (Figure~\ref{batlc-3}) consists of a single pulse of $\sim$0.2~s with a faint precursor emission (inset of Figure~\ref{batlc-3}) visible in the soft energy range ($<$50~ keV). XRT and UVOT observations started $\sim$2.3 days after the GRB and found no credible counterpart \citep{gcn080121}.

Two bright SDSS galaxies, SDSS J090858.15+414926.5 (G1 in Figure~\ref{batlc-3}) and SDSS J090904.12+415033.2 (G2 in Figure~\ref{batlc-3}) lie within the BAT position \citep{lien16}. The maximum projected offset, derived considering the maximum distance between the galaxy's center and the border of the 90\% error region, is  $\sim$220$\arcsec$ ($\sim$200 kpc) and $\sim$180$\arcsec$ ($\sim$170 kpc) from G1 and G2, respectively.

G1 is a face-on late-type galaxy at a distance of 203$\pm$14 Mpc with an 
absolute B-band magnitude $M_{B}$=-18.5 mag. 
G2 is instead an early-type galaxy at 207$\pm$14 Mpc with $M_{B}$=-18.8 mag. 
If the burst is indeed associated to any of these galaxies, 
its energy release would be $\approx1.6\times10^{47}$ erg in the 15-150 keV energy range. The precursor displays properties similar to the sample
presented in \citet{Troja10}, but, at this distance, would
have a much lower energy budget $\approx$3$\times10^{46}$\,erg. 

G1 is reported in the WISE catalogue \citep{cutri2013wise} 
with $W1(3.4 \mu m)$=18.94$\pm$0.05 AB mag and $W2(4.6 \mu m)$=19.43$\pm$0.15 AB mag, and in the GALEX catalogue with
m$_{FUV}$=19.83$\pm$0.12 AB mag, m$_{NUV}$=19.52$\pm$0.06 AB mag. 
Based on these values, we estimate a stellar mass log(M/M$_{\odot}$)$\approx$8.1 \citep{wen2013} and an unobscured star formation rate of log(SFR/M$_{\odot}$ yr$^{-1})$ $\approx$ -0.27 \citep{iglesias2006}. 
G2 displays a substantially redder color, with $W1(3.4 \mu m)$=16.80$\pm$0.03 AB mag, $W2(4.6 \mu m)$=17.36$\pm$0.04 AB mag, and m$_{NUV}$=22.00$\pm$0.52 AB mag. 
This suggests a larger stellar mass, log(M/M$_{\odot}$)$\approx$9.0 \citep{wen2013}, 

and a lower star formation rate, log(SFR/M$_{\odot}$ yr$^{-1})$ $\lesssim$ -0.35 \citep{iglesias2006}. 
Considered the heterogeneous environment of short GRBs, both these galaxies are plausible host. 

This burst occurred between LIGO S5 and S6 runs, and no constraints on its
distance are available. 

 \subsubsection{GRB100216A}

GRB 100216A was discovered through the ground-based analysis of {\it Swift}/BAT data \citep{cummings10}, and consists of a single pulse of duration $\sim$0.2 s (Figure~\ref{batlc-4}). 
The signal significance in BAT is marginal \citep{lien16}, however the same event was detected by the {\it Fermi} Gamma-Ray Burst Monitor, which confirms it as a real burst. XRT and UVOT observations started $\sim$2.5 days after the GRB and found no credible counterpart \citep{gcn100216a}.

A bright SDSS galaxy at redshift $z$=0.038 ($\sim$172~Mpc) lies within the BAT error region. The maximum projected offset from the galaxy  is $\sim$220$\arcsec$, corresponding to
$\sim$180 kpc.

At this distance the GRB isotropic-equivalent energy release would be $E_{\gamma,iso} \approx 9\times 10^{46}$ erg (15-150 keV). 
The putative host galaxy, SDSS J101700.25+353118.9 (LEDA 86918),
appears as a face-on barred spiral, 
its IR luminosity from WISE (W1$(3.4 \mu m)$=18.03$\pm$0.04 AB mag, W2$(4.6 \mu m)$=18.57$\pm$0.09 AB mag) suggests a stellar mass of $log(M/M_{\odot})\approx$8.3 \citep{wen2013}. From its UV luminosity  (m$_{FUV}$=18.89$\pm$0.11 AB mag, m$_{NUV}$=18.69$\pm$0.08 AB mag) 
we derive an unobscured star formation rate of log(SFR/M$_{\odot}$ yr$^{-1})$ $\approx$ -0.12 \citep{iglesias2006}. 

This burst happened during LIGO's sixth science run. Analysis of the GW data excludes that the merger happened within 23 Mpc for a NS-NS merger, and 40 Mpc for a NS-BH merger (90\% confidence level, \citealt{Abadie2012}).

\begin{center}
\begin{figure*}
\includegraphics[scale=0.69]{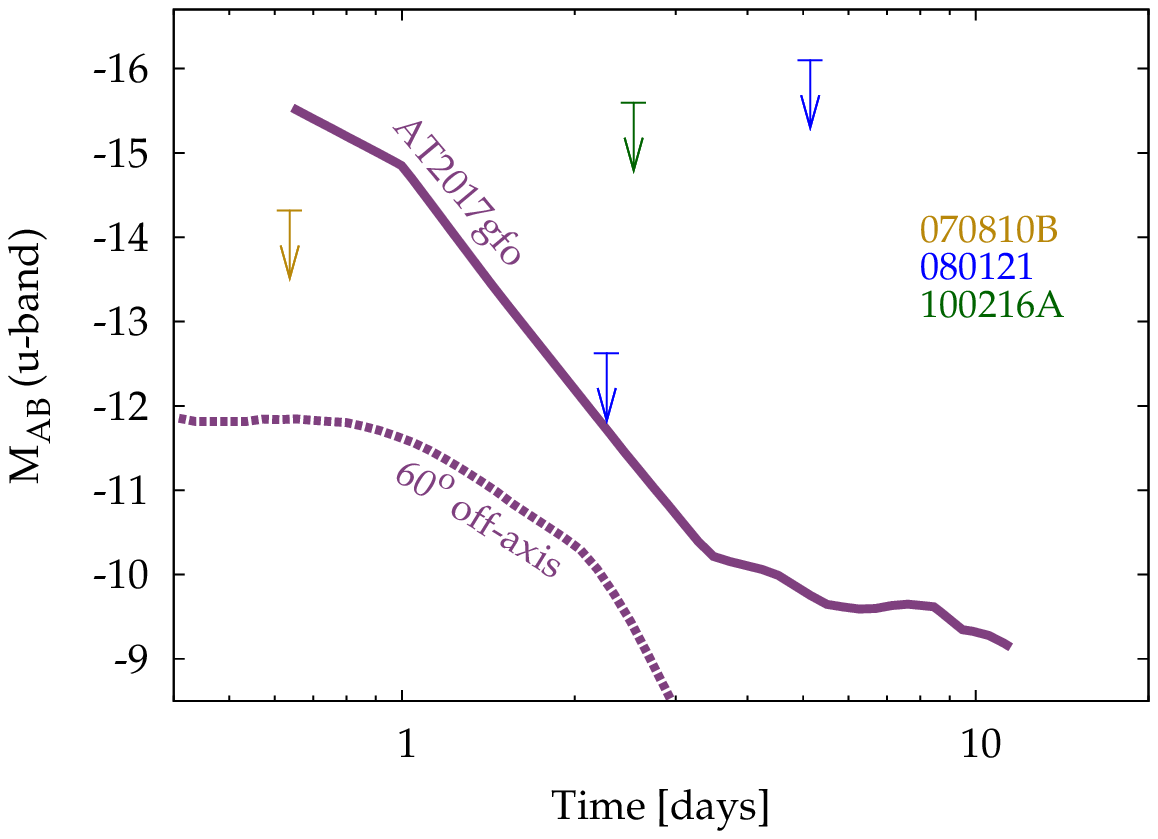}
\includegraphics[scale=0.69]{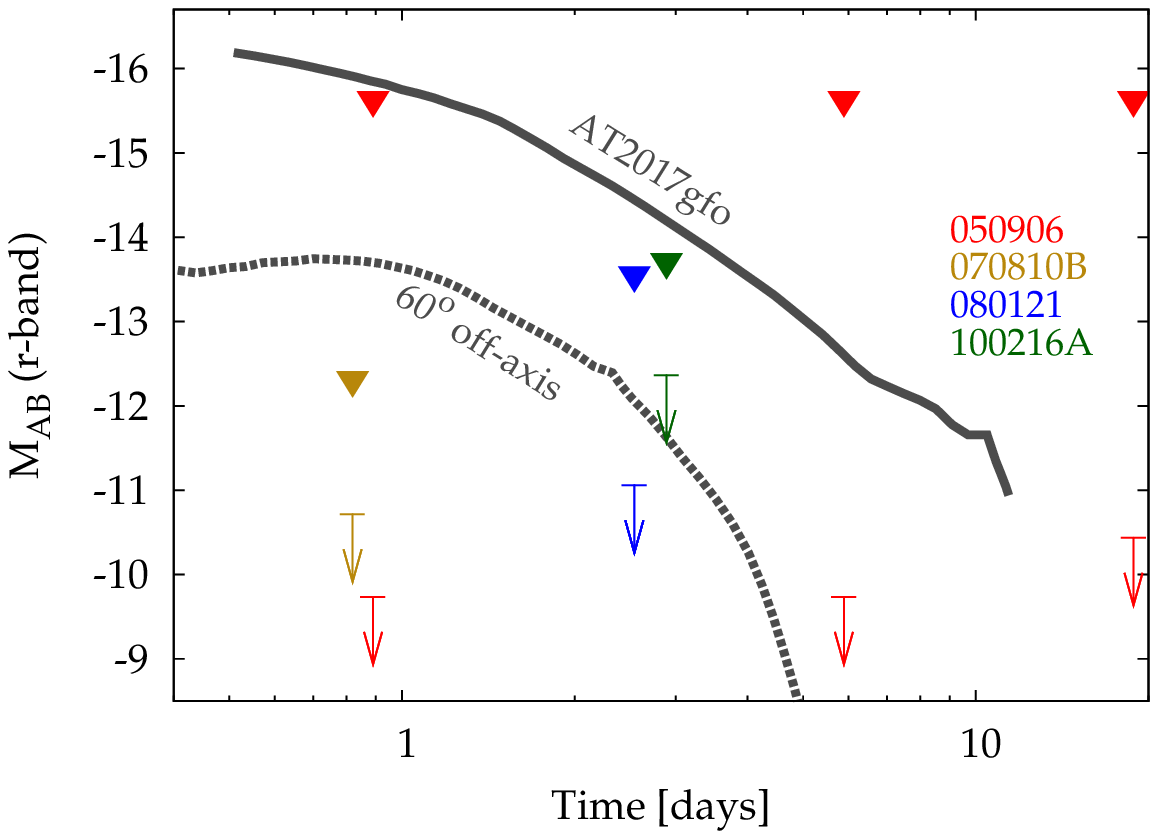}
\caption{
Optical upper limits for the four candidate nearby events in the UVOT u-band (left panel) and in the r-band (right panel). 
Solid lines show the temporal evolution of the kilonova AT2017gfo in the two filters \citep{Rossi2019}. Dashed lines show the synthetic light curve for the two component (red and blue) kilonova observed with an angle of 60$^\circ$ from jet axis (from \citet{troja2017}).
Two types of optical upper limits are reported: those derived for field objects (downward arrows), and those for sources near the galaxy's center (downward triangles). Only field limits are derived for UVOT images.
\label{upper_limits}}
\end{figure*}
\end{center}

\subsubsection{Optical Limits}

In Figure~\ref{upper_limits} we compare the optical limits for the four candidate nearby bursts with the kilonova AT2017gfo.
In no case an optical counterpart was detected within or close to the putative host galaxy.
Limits from {\it Swift}/UVOT and ground-based optical imaging are reported in Figure~\ref{upper_limits}, and compared with the light curve of AT2017gfo.

At distances larger than 100 Mpc, the UVOT sensitivity is comparable to or
shallower than the predicted emission, and the presence of a kilonova 
cannot be meaningfully constrained (Figure~\ref{upper_limits}; left panel). 
Deep ground-based imaging yields much
tighter limits (Figure~\ref{upper_limits}; right panel), although the sensitivity of the search is affected by the galaxy's light. For this reason, in Figure~\ref{upper_limits} (right panel) we report two sets of upper limits, derived within the galaxy (downward triangles) and outside it (downward arrows).
The internal limits are derived using image subtraction technique and simulating point-like sources close to the center of the galaxy. All the limits are reported in an online table (supplementary material).  
For GRB~100216A only a single epoch of observations is available, 
and upper limits were derived by using  reference images from PanSTARRs \citep{Chambers16}.
The higher background level and image subtraction artifacts in the central regions of the galaxy might prevent a reliable source detection. 
The contours of these regions are shown in Figures~\ref{batlc-2}, ~\ref{batlc-3}, and ~\ref{batlc-4}. 
Fake sources were simulated immediately out of these regions to derive the internal limits.
Based on the offset distributions of sGRBs \citep[e.g.][]{berger14}, the probability to find a sGRB at similar or smaller offsets from the galaxy's center
is 18\%, 13\% and 35\% for GRB 070810B, GRB 080121 and GRB 100216A, respectively.

For events outside the red contours (Figures~\ref{batlc-2}, ~\ref{batlc-3}, and ~\ref{batlc-4}), the presence of a kilonova similar to AT2017gfo can be ruled out in the optical. 
It's worth to note that infrared observations are available only for GRB 050906 \citep[see ][]{levan08} and the derived upper limits are 2.5 times higher than the magnitude expected for AT2017gfo.

We explored the consistency of a kilonova non-detection with these optical limits in terms of the key physical parameters $M_\textrm{ej}$ and $v_\textrm{ej}$, which are the ejecta mass and velocity respectively. We applied the simple analytic model outlined in \citet{hotokezaka2018} for the bolometric lightcurve, and set an opacity of $\kappa=0.1$ cm$^2$/g for the blue (lanthanide-poor) kilonova component and $\kappa=10$ cm$^2$/g for the red (lanthanide-rich) component \citep[e.g.][]{Roberts11,Grossman2014}. We convert the bolometric lightcurve as a function of time to an r-band magnitude assuming a blackbody spectrum with the temperature evolution outlined in \citet{hotokezaka2018}. 
Field upper limits place the tightest constraints and, for a velocity $v_\textrm{ej} \gtrsim 0.2$c, imply
an ejecta mass $M_\textrm{ej} \lesssim 10^{-3} M_\odot$ for events viewed towards their polar regions. 
This value is smaller than the one derived for GW170817A/AT2017gfo \citep{hotokezaka2018}
and other candidate kilonovae \citep{Troja2018a,Lamb19b,Troja2019b}.

The typical parameters for ejecta mass and velocity in compact mergers is found to be $M_\textrm{ej} =10^{-3}-10^{-1} M_\odot$, with typical velocities of $v_\textrm{ej}=0.1-0.3$c \citep[e.g.]{Bauswein13,Hotokezaka13,Perego14}. Thus, our field limits are quite restrictive on the typical parameter space for r-process ejecta, and lead us to conclude that a lanthanide-poor kilonova viewed towards the polar regions would likely have been detected.
For events closer to the galaxy's center 
the limits are shallower, and ejecta
masses $M_\textrm{ej} \lesssim 10^{-2} M_\odot$ could be compatible with the observations. 

For an off-axis observer, the emission is suppressed by the lanthanide-rich ejecta and can be more than two magnitudes lower than AT2017gfo. In Figure~\ref{upper_limits}, we show the predicted kilonova emission for an observer located 60$^{\circ}$ off-axis (dashed lines) as presented by \citet{troja2017} and derived using the model of \citet{Wollaeger2018}. In this case only the optical limits from GRB 070810B and the field limits from GRB~050906 remain constraining. 

The lack of an optical counterpart 
can also constrain the presence of a lanthanide-rich kilonova. Our limits imply a low mass of neutron-rich ejecta, $M_\textrm{ej}<10^{-2.1} M_{\odot}$, 
for any event in the galaxy's outskirts. However, if we consider the shallower upper limits derived in the inner galaxy's regions, then a lanthanide-rich kilonova similar to AT2017gfo could be consistent with our lack of detection. 

\subsection{Comparison with previous works} 

Our results present some differences with the works of \citet{mandhai18}
and \citet{bartos19}, which addressed a similar topic. 
\citet{mandhai18} considered a sample of 150 {\it Swift} GRBs with $T_{90}$<4~s, longer than the duration threshold used in this work,
and included both GRBs with a detected afterglow and those without any counterpart. 
They adopted a different galaxy catalogue \citep[2MASS Redshift Survey; ][]{Huchra2012} and a visual examination of DSS II images.
Their matching radius for nearby galaxy extended to a distance of 200 kpc from the GRB position. 
This value is much larger than the typical projected offset of short GRBs \citep{Troja2008,Fong13,Tunnicliffe2014} and, whereas it does not significantly increase the chance of finding a real association, it more than doubles the probability of chance alignments, from 3\% of our sample to 8\%. 
Using this approach the authors find an upper limit on the all-sky rate of local short GRB detectable by {\it Swift} of <$4$ yr$^{-1}$.  

We conclude that differences in the adopted catalogues of galaxies
as well as in the selected GRB sample result in a different list of candidates and a different estimate for all-sky rate. Looser constraints on the burst duration and maximum GRB/galaxy offset increased the probability of spurious associations. 

Our selection criteria and search strategy are more similar to those
presented in \citet{bartos19}. Their work focuses on late-time radio monitoring of sGRBs, and thus excludes all the sources with declination below -40$^{\circ}$ and between -5$^{\circ}$ and 15$^{\circ}$ due to satellites interference. 
They also use optical upper limits from the literature to constrain any kilonova signal, and conclude that $\sim1/3$ of sGRBs in their sample can not be associated with a AT2017gfo-like kilonova if located at $\lesssim$200~Mpc. However, they do not consider viewing angle effects 
nor the contamination from the nearby bright galaxy, and how it affects optical searches. 
\citet{bartos19} adopt a larger search radius for local galaxies 
and derive a higher chance probability for the galaxy association. While in our work we derive the chance probability from the the all sky distribution of galaxies, \citet{bartos19} derive it from the density of galaxies around the GRB position. 
Due to inhomogeneities in the local universe \citep[e.g.][]{sylosL2011},
a nearby galaxy may reside within an overdensity of galaxies, thus the chance probability as calculated by \citet{bartos19} could be high also in the case of a true local sGRB. 

Another fundamental difference between our work and past searches is that we took into account the observing conditions, removing from our sample
bursts affected by sub-optimal observations. 

\section{Discussion} \label{sec:discussion}
 \subsection{Implications for the rate of NS mergers}

\begin{center}
\begin{figure}
\includegraphics[scale=0.74]{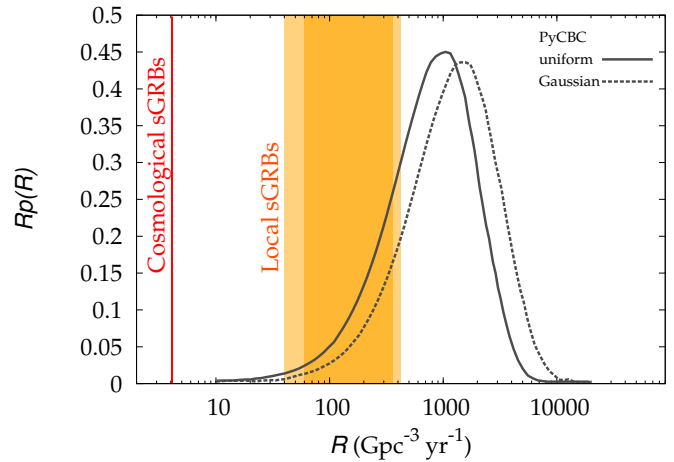}
\caption{Isotropic equivalent rate density for sGRBs.
Dark (light) orange region shows the 68\% (90\%) confidence interval for the rate of local sGRBs. 
The vertical solid line shows the rate of cosmological sGRBs derived by \citet{wp15}.
These results are compared with the posterior probability distribution of binary neutron star mergers derived by the LIGO/VIRGO collaboration (\citealt{GWTC-1}; PyCBC matched-filter search).
Dashed and dotted lines are derived assuming a uniform and Gaussian mass distribution, respectively.  
 } 
\label{rate}
\end{figure}
\end{center}
Our search did not find any event closer than 40 Mpc, in agreement with 
previous limits derived from joint GW-GRB searches \citep{Abadie2010,Abadie2012}.
Since galaxy catalogues are complete within this distance range, we conclude that no GRB discovered by {\it Swift} was as close as GW170817. 
We also did not find any event in the range 40 - 100 Mpc, although the GLADE galaxy catalogue is not complete in this distance range \citep{glade}. 
Four plausible candidates were found in the range 100-200 Mpc, which, 
given the larger volume included, is consistent with the lack of detection at closer distances.

Two of the candidate bursts were discovered by a human-based analysis, which was neither systematic nor homogeneous.  
For this reason, we do not use them to calculate the rate of events. 
The other two bursts (GRB~050906 and GRB~070810B) were discovered by the standard trigger algorithm, 
which can be modeled in order to estimate the detectability of these objects \citep[e.g.][]{Lien2014}. 
The corresponding all-sky (isotropic equivalent) rate of events (together with the corresponding 1$\sigma$ confidence levels) was derived assuming a local origin for this subsample of sGRBs: 

\begin{equation}\label{eq:rate}
\mathfrak{R} = \frac{1}{T} \frac{4 \pi}{\Omega_{max}} f_G \sum_i \frac{1}{V_{i,\rm max}} = 160^{+200}_{-100}  {\rm Gpc}^{-3}  {\rm yr}^{-1}
\end{equation}
Assuming that all the binary NS mergers result in successful GRB jets, this rate implies a beaming factor
$f_b^{-1}=10^{+70}_{-8}$ (90\% confidence interval)
to be consistent with the NS merger rate derived by the LIGO-Virgo collaboration 
\citep{GWTC-1}.
\textit{V}$_{i,\rm max}$ in Equation (\ref{eq:rate}) is the maximum volume of detectability for each event $i$,
calculated considering the variation of the observable solid angle $\Omega$ with respect to the distance $r$ of the burst, \textit{V}$_{i,\rm max}$= $\Omega_{max}^{-1} \int $\textit{V}$(r) \Omega(r)$ d$r$. We used the sensitivity curves presented in \citet{Lien2014} to parameterize this effect. 
The maximum solid angle, $\Omega_{max}\approx2$\,sr, corresponds to a partial coding $>10\%$, and the control time $T\approx11$\,yr was estimated for an average {\it Swift} duty cycle of 78\%.
The factor $f_G$ takes into account the completeness of the galaxy catalogue. Following that sGRBs preferentially occur in the brightest galaxies, the GLADE catalogue 
can be considered $\sim$90\% complete up to 200 Mpc \citep{glade}, 
and $f_G\approx$1.1.
We caution that the result is mostly driven by the weakest event, GRB 050906, and the volumetric rate derived only from GRB~0708010B would be $\approx$30~Gpc$^{-3}$~yr$^{-1}$, a factor of five lower. 

We can note from Figure~\ref{rate} that this rate is significantly higher that the one reported from measurements derived using cosmological GRBs (e.g. $\approx$4 Gpc$^{-3}$ yr$^{-1}$ from  \citealt[][]{wp15}). This discrepancy 
is likely due to the higher number of low-luminosity sGRBs detectable in the local Universe. Indeed, in their calculations \cite{wp15} imposed a minimum isotropic-equivalent luminosity of $5 \times 10^{49}$ erg s$^{-1}$, whereas our results are driven by two under-luminous events. Extrapolating the results of \cite{wp15} to lower values of minimum luminosity, $L_{min} \approx 10^{47}$, we found a rate consistent with the one presented in Equation (\ref{eq:rate}).

Based on the above numbers and on the assumption that our candidates reside within 200 Mpc, the local all-sky rate of detectable sGRBs is $1.3^{+1.7}_{-0.8}$  yr$^{-1}$ (68\% confidence interval). By taking into account the instruments'
field of view and duty cycle, this corresponds to
$0.16^{+0.2}_{-0.10}$  events yr$^{-1}$ in the triggered {\it Swift} sample, and $0.8^{+1.0}_{-0.5}$ events yr$^{-1}$ in the {\it Fermi} sample, due the larger field of view of the Gamma-Ray Burst Monitor. 
We also find that a systematic search of the untriggered {\it Swift} dataset could increase the rate of joint GRB/GW detections, and the chances of rapid and accurate localization of a GW source.

If the identified candidates reside instead at higher redshift, we derive
an upper limit of  $\lesssim$2 events yr$^{-1}$ (90\% confidence interval) to the rate of nearby detectable sGRBs.
During the selection process we did not account for the subsample of sGRBs with extended emission, however this should not substantially affect our results. 
These bursts represent $\lesssim$15\% of the {\it Swift} sGRB sample and, if included, 
they may
increase the derived event rate by a comparable fraction.

Our results show that the number of detectable sGRBs within 200 Mpc could be substantially higher than pre-GW predictions \citep[e.g.][]{clark15, ghirlanda2016}, and are consistent with independent estimates based on the {\it Fermi} dataset \citep[e.g.][]{LVCGBM,burgess17,zhang18}.
Our value is, in all cases, smaller than the optimistic estimates by \citet{GupteBartos}, and in line with the constraints derived by \citet{mandhai18}.

\begin{table}
 	\centering
 	\caption{Results of Monte Carlo simulations reporting the beaming factor $f_{b}^{-1}$ for different jet structures. We used the broken power-law models for gamma-ray luminosity from \citet{wp15} (WP15) and \citet{ghirlanda2016} (G16). 
 	}
 	\label{tab:beamingfactor}
 	\begin{tabular}{lcc}
    \hline
    Jet Model \hspace{3cm} & WP15 & G16 \\
    \hline
    Top-Hat ($\theta_j=0.1$) \dotfill & 200 & 200 \\
    
    Gaussian ($\theta_{core}=0.1$) \dotfill & 33 & 27 \\
    PL  ($\delta=4$) \dotfill & 26 & 4 \\
    PL ($\delta=6$)  \dotfill & 54 & 14 \\
    Cocoon-like ($\eta_\textrm{br}=10^{-3}$) \dotfill & 72 & 29 \\
    Cocoon-like  ($\eta_\textrm{br}=10^{-2}$) \dotfill & 5 & 1.5 \\
    \hline
    \end{tabular}
\end{table}

\subsection{Constraints on the jet structure}

The conversion of the observed rate to the intrinsic rate typically involves a geometrical beaming fraction $f_b$, which for a top-hat jet is $f_b=1-\cos\theta_j\approx \theta_j^2/2$, where $\theta_j$ is the jet half-opening angle.
Interestingly the derived rate of visible sGRBs is already close to the NS merger rate, 
110$<\mathfrak{R}<$3840 Gpc$^{-3}$ yr$^{-1}$  
(90\% confidence interval),
as  estimated from GW data
\citep{GWTC-1}.
This suggests a low beaming factor
$f_b^{-1}=10^{+70}_{-8}$ (90\% confidence interval),
as well as an efficient production of sGRBs, i.e. most neutron stars mergers may result in successful GRB jets \citep[see also ][]{beniamini2019b,Beniamini2020}. These results are consistent also with the most updated estimates of the NS rate obtained during the O3 run of Advanced LIGO and Advanced Virgo \citep{GW190425}.

In cosmological sGRBs the beaming factor is commonly estimated from the afterglow jet-breaks, and found to be of the order of $f_b^{-1}\approx$200 \citep[e.g.][]{Burrows06,Troja16,Jin18}. 

However, recent observations of GW170817 and its afterglow revealed a complex structure of the relativistic outflow. In particular, whereas the common belief was that the gamma-ray emission is visible only to observers located within the narrow jet core, observations of GW170817 suggested that observers located at large angles from the jet-axis can still detect a faint gamma-ray signal. These effects are insignificant at cosmological distances \citep{beniamini2019a}, but should be taken into account in the nearby Universe, where this faint prompt emission becomes detectable by current gamma-ray facilities. 
For GW170817, afterglow modeling finds that the ratio between the viewing angle $\theta_{view}$
and the jet core opening angle $\theta_{core}$ is $\theta_{view}/\theta_{core} \approx$ 5-6,
with $\theta_{core} \approx$5~deg \citep[e.g.][]{Mooley18,Troja2019a,Lamb19}. 
The resulting beaming correction is only $f_b^{-1}\approx$10. 

We performed a series of Monte Carlo simulations to further explore the consistency of different angular and radial jet profiles with the rate of local sGRBs identified in this work. 
We explored a variety of jet structures in order to identify which structures have geometrical beaming fractions $f_b^{-1} \lesssim$80, so that the intrinsic local rate of sGRBs is consistent with the rate of NS mergers. The jet structures explored are (i) a power-law (PL) energy dependence beyond the core, (ii) a Gaussian function in angle from the core, and (iii) a `cocoon'-like model which involves a quasi-isotropic weak component surrounding the jet's core (see e.g. \citealt{Ryan2019} and \citealt{beniamini2019b} for the details of these jet structures).
For each burst in the simulation, we sampled a random orientation of the jet relative to the line of sight ($\theta_{obs}$), 
and a distance $d$, according to the volume of the local universe. The energy of the jet's core is simulated according to the luminosity function of sGRBs from \citet{wp15} and \citet{ghirlanda2016} (case $a$). 
We convert the gamma-ray luminosity to gamma-ray energy assuming a typical rest frame duration $\langle T_{90}\rangle=0.2$ s, $E_{\gamma,\textrm{iso}} \approx \langle T_{90}\rangle L_{\gamma,\textrm{iso}}$. 
The observed energy $E(\theta_{obs})$ is then computed according to the jet's profile, and converted into a fluence using average spectral parameters for sGRBs \citep{lien16}. 
We estimate the fraction of sGRBs detectable by \textit{Swift} by applying an approximate minimum detection threshold on the observed fluence $F_{\gamma,\textrm{lim}}\approx2\times10^{-8}$ erg/cm$^{2}$ in the 15-150 keV band \citep[from Figure 8 in ][]{Lien2014}, which allows us to roughly estimate the beaming factor for each model.  
For each jet structure, the effective beaming factor $f_b$ is calculated as the ratio
between the number of detected bursts and the total number of simulations
($N=2\times10^5$).

The results are tabulated in Table~\ref{tab:beamingfactor}. 
We find that there are a variety of structured jet models consistent with $f_b^{-1}\lesssim 80$: both a Gaussian
profile and a power-law profile with 
slope $\delta\approx5$ fit the observations. Cocoon-like models require an efficient break-out, $10^{-3}<\eta_\textrm{br} < 10^{-2}$, in order 
to reproduce the observed rates. 
These conclusions are not particularly sensitive to the choice of luminosity function.

\section{Conclusion}\label{sec:conclusions}

We examined the {\it Swift} database searching for low-luminosity sGRBs in the local Universe, analogous to GW170817/GRB170817A. Despite their close distance these events were not discovered before
the advent of GW astronomy. We found that only a small fraction ($\lesssim$5\%) 
of {\it Swift} short GRBs could potentially be located within 200 Mpc, and that follow-up observations were not sufficient to constrain their nature. A combination of low number statistics and sub-optimal observing strategy could explain the lack of identification. 

Assuming a local origin for this subsample of sGRBs, we find an all-sky (isotropic equivalent) rate density of $160^{+200}_{-100}$ ${\rm Gpc}^{-3}  {\rm yr}^{-1}$. If all the binary NS mergers result in successful GRB jets, this rate implies a beaming factor
$f_b^{-1}=10^{+70}_{-8}$ 
to be consistent with the NS merger rate derived by the LIGO-Virgo collaboration.
This result allows us to disfavor top-hat jet models and cocoon-like models with  inefficient breakout. Different configurations of structured jet models (such as Gaussian or power-law models) are consistent with the observational constraints.

By using the upper limits placed in the optical band, we also provide constraints on the possible kilonova emission, and the allowed mass and velocity of the merger ejecta. 
A lanthanide-poor kilonova viewed towards the polar regions would have likely been detected by ground-based optical observations, but no strong constraints can be placed for off-axis events or lanthanide-rich kilonovae.

We cannot exclude the possibility that none of the reported candidates occurred within 200 Mpc. 
For typical sGRBs parameters and cosmological distances, the lack of X-ray afterglow could be explained by a tenuous environment.
In this case, the upper limit on the rate of local events would be $\lesssim180 {\rm~Gpc}^{-3}  {\rm yr}^{-1}$ (90\% confidence level). 
Given the rate of binary NS mergers predicted by LIGO, a minimum beaming factor of $f_b^{-1} \gtrsim$10 would be consistent with no detection of local sGRBs by {\it Swift}.

An optimization of the follow-up strategies 
and a systematic search for untriggered bursts could be crucial to increase the detection rate of local events.   

\section*{Acknowledgements}
We thank Brad Cenko and Antonio Galvan-Gamez for the help with data analysis and the useful discussions. 
This work was supported in part by the National Aeronautics and Space Administration through grant NNX10AF62G issued through the Astrophysics Data Analysis Program. 
This work was performed in part at Aspen Center for Physics, which is supported by National Science Foundation grant PHY-1607611.
The research of PB was funded by the Gordon and Betty Moore Foundation through Grant GBMF5076.
This research uses services or data provided by the Science Data Archive at NOAO. NOAO is operated by the Association of Universities for Research in Astronomy (AURA), Inc. under a cooperative agreement with the National Science Foundation.
This publication made use of data products supplied by the UK Swift Science Data Centre at the University of Leicester, 
and from the Wide-field Infrared Survey Explorer, which is a joint project of the University of California, Los Angeles, and the Jet Propulsion Laboratory/California Institute of Technology, funded by the National Aeronautics and Space Administration.


\label{lastpage}


\begin{thebibliography}{99}

\bibitem[Abadie et al.(2010)]{Abadie2010} Abadie, J., Abbott, B.~P., Abbott, R., et al.\ 2010, \apj, 715, 1453.

\bibitem[Abadie et al.(2012)]{Abadie2012} Abadie, J., Abbott, B.~P., Abbott, R., et al.\ 2012, \apj, 760, 12.

\bibitem[Abbott et al.(2017a)]{LVCGBM} Abbott, B.~P., Abbott, R., Abbott, T.~D., et al.\ 2017, \apjl, 848, L13 

\bibitem[Abbott et al.(2017b)]{Abbot17m} Abbott, B.~P., Abbott, R., Abbott, T.~D., et al.\ 2017, \apjl, 848, L12.

\bibitem[Abbott et al.(2019)]{GWTC-1} Abbott, B.~P., Abbott, R., Abbott, T.~D., et al.\ 2019, Physical Review X, 9, 031040.

\bibitem[Bartos et al.(2019)]{bartos19} Bartos, I., Lee, K.~H., Corsi, A., M{\'a}rka, Z., \& M{\'a}rka, S.\ 2019, \mnras, 485, 4150

\bibitem[Bauswein, Goriely, \& Janka(2013)]{Bauswein13} Bauswein, A., Goriely, S., \& Janka, H.-T.\ 2013, \apj, 773, 78.

\bibitem[Becker(2015)]{Becker2015} Becker, A.\ 2015, Astrophysics Source Code Library, ascl:1504.004.

\bibitem[\protect\citeauthoryear{Beniamini et al.}{2015}]{Beniamini2015} Beniamini P., Nava L., Duran R.~B., Piran T., 2015, MNRAS, 454, 1073

\bibitem[\protect\citeauthoryear{Beniamini \& van der Horst}{2017}]{BeniaminiVanDerHorst} Beniamini P., van der Horst A.~J., 2017, MNRAS, 472, 3161

\bibitem[Beniamini \& Nakar(2019)]{beniamini2019a} Beniamini, P., \& Nakar, E.\ 2019, \mnras, 482, 5430 

\bibitem[\protect\citeauthoryear{Beniamini et al.}{2019}]{beniamini2019b} Beniamini P., Petropoulou M., Barniol Duran R., Giannios D., 2019, MNRAS, 483, 840

\bibitem[\protect\citeauthoryear{Beniamini, et al.}{2020}]{Beniamini2020} Beniamini P., Barniol Duran R., Petropoulou M., Giannios D., 2020, arXiv, arXiv:2001.00950

\bibitem[Berger(2014)]{berger14} Berger, E.\ 2014, \araa, 52, 43 

\bibitem[Bianchi et al.(2017)]{Bianchi17} Bianchi, L., Shiao, B., \& Thilker, D.\ 2017, \apjs, 230, 24 

\bibitem[Blanton et al.(2017)]{Blanton2017} Blanton, M.~R., Bershady, M.~A., Abolfathi, B., et al.\ 2017, \aj, 154, 28 

\bibitem[Breeveld et al.(2010)]{Breeveld2010} Breeveld, A.~A., Curran, P.~A., Hoversten, E.~A., et al.\ 2010, \mnras, 406, 1687.

\bibitem[Bromberg et al.(2013)]{Bromberg13} Bromberg, O., Nakar, E., Piran, T., \& Sari, R.\ 2013, \apj, 764, 179 

\bibitem[Burgess et al.(2017)]{burgess17} Burgess, J.~M., Greiner, J., Begue, D., et al.\ 2017, arXiv e-prints, arXiv:1710.05823 

\bibitem[Burrows et al.(2006)]{Burrows06} Burrows, D.~N., Grupe, D., Capalbi, M., et al.\ 2006, \apj, 653, 468.

\bibitem[Chambers et al.(2016)]{Chambers16} Chambers, K.~C., Magnier, E.~A., Metcalfe, N., et al.\ 2016, arXiv e-prints, arXiv:1612.05560.

\bibitem[\protect\citeauthoryear{Clark, et al.}{2015}]{clark15} Clark J., et al., 2015, \apj, 809, 53

\bibitem[\protect\citeauthoryear{Coulter, et al.}{2017}]{Coulter17} Coulter, D.~A., Foley, R.~J., Kilpatrick, C.~D., et al.\ 2017, Science, 358, 1556.

\bibitem[Coward et al.(2012)]{cow12} Coward, D.~M., Howell, E.~J., Piran, T., et al.\ 2012, \mnras, 425, 2668 

\bibitem[Cummings \& Palmer(2008)]{gcn7209} Cummings, J.~R., \& Palmer, D.~M.\ 2008, GRB Coordinates Network, 7209, 1.

\bibitem[Cummings et al.(2010)]{cummings10} Cummings, J.~R., Barthelmy, S.~D., Fox, D.~B., et al.\ 2010, GRB Coordinates Network, Circular Service, No.~10428, \#1 (2010), 10428, 1

\bibitem[Cummings et al.(2010)]{gcn11436} Cummings, J.~R., Palmer, D.~M., Barthelmy, S.~D., et al.\ 2010, GRB Coordinates Network, 11436, 1.

\bibitem[Cutri et al.(2013)]{cutri2013wise} Cutri, R.~M., Wright, E.~L., Conrow, T., et al.\ 2013, Explanatory Supplement to the AllWISE Data Release Products, 1.

\bibitem[D{\'a}lya et al.(2018)]{glade} D{\'a}lya, G., Galg{\'o}czi, G., Dobos, L., et al.\ 2018, \mnras, 479, 2374 

\bibitem[D'Avanzo et al.(2018)]{Davanzo18} D'Avanzo, P., Campana, S., Salafia, O.~S., et al.\ 2018, \aap, 613, L1.

\bibitem[Eichler et al.(1989)]{eichler89} Eichler, D., Livio, M., Piran, T., \& Schramm, D.~N.\ 1989, \nat, 340, 126 

\bibitem[Evans et al.(2009)]{evans09} Evans, P.~A., Beardmore, A.~P., Page, K.~L., et al.\ 2009, \mnras, 397, 1177

\bibitem[\protect\citeauthoryear{Fong \& Berger}{2013}]{Fong13} Fong W., Berger E., 2013, ApJ, 776, 18

\bibitem[Faber et al. (2006)]{faber06} Faber, J.~A., Baumgarte, T.~W., Shapiro, S.~L., \& Taniguchi, K.\ 2006, \apjl, 641, L93.

\bibitem[Gao et al.(2013)]{Gao2013} Gao, H., Ding, X., Wu, X.-F., Zhang, B., \& Dai, Z.-G.\ 2013, \apj, 771, 86.

\bibitem[Gehrels(1986)]{Gehrels1986} Gehrels, N.\ 1986, \apj, 303, 336.

\bibitem[Gehrels et al.(2004)]{swift04} Gehrels, N., Chincarini, G., Giommi, P., et al.\ 2004, \apj, 611, 1005 

\bibitem[Gehrels et al.(2005)]{gehrels05} Gehrels, N., Sarazin, C.~L., O'Brien, P.~T., et al.\ 2005, \nat, 437, 851 

\bibitem[Gehrels et al.(2016)]{Gehrels2016} Gehrels, N., Cannizzo, J.~K., Kanner, J., et al.\ 2016, \apj, 820, 136.

\bibitem[\protect\citeauthoryear{Ghirlanda et al.}{2016}]{ghirlanda2016} Ghirlanda G., et al., 2016, A\&A, 594, A84
\bibitem[\protect\citeauthoryear{Ghirlanda, et al.}{2018}]{Ghirlanda2017} Ghirlanda G., et al., 2018, A\&A, 609, A112

\bibitem[Giacomazzo et al.(2013)]{jack13} Giacomazzo, B., Perna, R., Rezzolla, L., Troja, E., \& Lazzati, D.\ 2013, \apjl, 762, L18 


\bibitem[Gompertz et al.(2018)]{gomp18} Gompertz, B.~P., Levan, A.~J., Tanvir, N.~R., et al.\ 2018, \apj, 860, 62 

\bibitem[\protect\citeauthoryear{Granot \& Sari}{2002}]{Granot2002} Granot J., Sari R., 2002, ApJ, 568, 820
\bibitem[\protect\citeauthoryear{Grossman et al.}{2014}]{Grossman2014} Grossman D., Korobkin O., Rosswog S., Piran T., 2014, MNRAS, 439, 757

\bibitem[\protect\citeauthoryear{Gupte \& Bartos}{2018}]{GupteBartos} Gupte N., Bartos I., 2018, arXiv, arXiv:1808.06238

\bibitem[Hallinan et al.(2017)]{Hallinan2017} Hallinan, G., Corsi, A., Mooley, K.~P., et al.\ 2017, Science, 358, 1579.

\bibitem[Hook et al.(2004)]{hook2004} Hook, I.~M., J{\o}rgensen, I., Allington-Smith, J.~R., et al.\ 2004, \pasp, 116, 425.

\bibitem[Hotokezaka et al.(2013)]{Hotokezaka13} Hotokezaka, K., Kiuchi, K., Kyutoku, K., et al.\ 2013, \prd, 87, 024001.

\bibitem[\protect\citeauthoryear{Hotokezaka et al.}{2018}]{hotokezaka2018} Hotokezaka K., Beniamini P., Piran T., 2018, IJMPD, 27, 1842005

\bibitem[Huchra et al.(2012)]{Huchra2012} Huchra, J.~P., Macri, L.~M., Masters, K.~L., et al.\ 2012, \apjs, 199, 26.

\bibitem[Iglesias-P{\'a}ramo et al.(2006)]{iglesias2006} Iglesias-P{\'a}ramo, J., Buat, V., Takeuchi, T.~T., et al.\ 2006, \apjs, 164, 38

\bibitem[\protect\citeauthoryear{Im, et al.}{2017}]{Im17} Im M., et al., 2017, ApJL, 849, L16

\bibitem[Jin et al.(2015)]{jin15} Jin, Z.-P., Li, X., Cano, Z., et al.\ 2015, \apjl, 811, L22

\bibitem[Jin et al.(2018)]{Jin18} Jin, Z.-P., Li, X., Wang, H., et al.\ 2018, \apj, 857, 128.

\bibitem[Jin et al.(2019)]{Jin2019} Jin, Z.-P., Covino, S., Liao, N.-H., et al.\ 2019, Nature Astronomy, 461.
 
\bibitem[Kawaguchi et al.(2016)]{kawa16} Kawaguchi, K., Kyutoku, K., Shibata, M., \& Tanaka, M.\ 2016, \apj, 825, 52 

\bibitem[Kisaka \& Nakamura(2015)]{Kisaka2015} Kisaka, S., Ioka, K., \& Nakamura, T.\ 2015, \apj, 809, L8.

\bibitem[Kouveliotou et al.(1993)]{ck93} Kouveliotou, C., Meegan, C.~A., Fishman, G.~J., et al.\ 1993, \apjl, 413, L101 

\bibitem[Lamb et al.(2019)]{Lamb19} Lamb, G.~P., Lyman, J.~D., Levan, A.~J., et al.\ 2019, \apjl, 870, L15.

\bibitem[Lamb et al.(2019)]{Lamb19b} Lamb, G.~P., Tanvir, N.~R., Levan, A.~J., et al.\ 2019, \apj, 883, 48.

\bibitem[\protect\citeauthoryear{Lebrun, et al.}{2003}]{Lebrun03} Lebrun F., et al., 2003, Astronomy and Astrophysics, 411, L141

\bibitem[Levan et al.(2008)]{levan08} Levan, A.~J., Tanvir, N.~R., Jakobsson, P., et al.\ 2008, \mnras, 384, 541 

\bibitem[Lien et al.(2016)]{lien16} Lien, A., Sakamoto, T., Barthelmy, S.~D., et al.\ 2016, \apj, 829, 7 

\bibitem[Lien et al.(2014)]{Lien2014} Lien, A., Sakamoto, T., Gehrels, N., et al.\ 2014, \apj, 783, 24.

\bibitem[Mandhai et al.(2018)]{mandhai18} Mandhai, S., Tanvir, N., Lamb, G., Levan, A., \& Tsang, D.\ 2018, Galaxies, 6, 130

\bibitem[\protect\citeauthoryear{Margutti, et al.}{2018}]{Margutti18} Margutti R., et al., 2018, ApJL, 856, L18

\bibitem[\protect\citeauthoryear{Marshall, et al.}{2007}]{gcn070810b} Marshall, F.~E., Barthelmy, S.~D., Brown, P.~J., et al.\ 2007, GCN Report, 81, 1.

\bibitem[\protect\citeauthoryear{M{\'e}sz{\'a}ros \& Rees}{1997}]{Meszaros1997} M{\'e}sz{\'a}ros P., Rees M.~J., 1997, \apj, 476, 232

\bibitem[Metzger \& Piro(2014)]{Metzger2014} Metzger, B.~D., \& Piro, A.~L.\ 2014, \mnras, 439, 3916.

\bibitem[Mooley et al.(2018)]{Mooley18} Mooley, K.~P., Deller, A.~T., Gottlieb, O., et al.\ 2018, Nature, 561, 355.

\bibitem[Norris \& Bonnell(2006)]{Norris06} Norris, J.~P., \& Bonnell, J.~T.\ 2006, \apj, 643, 266.

\bibitem[Oke et al.(1995)]{oke1995} Oke, J.~B., Cohen, J.~G., Carr, M., et al.\ 1995, \pasp, 107, 375.

\bibitem[Parsons et al.(2005)]{gcn3935} Parsons, A., Sarazin, C., Barbier, L., et al.\ 2005, GRB Coordinates Network, 3935, 1.

\bibitem[Perego et al.(2014)]{Perego14} Perego, A., Rosswog, S., Cabez{\'o}n, R.~M., et al.\ 2014, \mnras, 443, 3134.

\bibitem[Perley et al.(2010)]{perley10} Perley, D.~A., Meyers, J., Hsiao, E., et al.\ 2010, GRB Coordinates Network, Circular Service, No.~10429, \#1 (2010), 10429, 1 

\bibitem[Planck Collaboration et al.(2018)]{planck2018} Planck Collaboration, Aghanim, N., Akrami, Y., Ashdown, et al.\ 2018, arXiv e-prints, arXiv:1807.06209.

\bibitem[Resmi et al.(2018)]{Resmi18} Resmi, L., Schulze, S., Ishwara-Chandra, C.~H., et al.\ 2018, \apj, 867, 57.

\bibitem[Rezzolla et al.(2011)]{rezzolla11} Rezzolla, L., Giacomazzo, B., Baiotti, L., et al.\ 2011, \apjl, 732, L6 

\bibitem[\protect\citeauthoryear{Roberts, et al.}{2011}]{Roberts11} Roberts L.~F., Kasen D., Lee W.~H., Ramirez-Ruiz E., 2011, ApJL, 736, L21


\bibitem[Roming et al.(2005)]{roming2005} Roming, P.~W.~A., Kennedy, T.~E., Mason, K.~O., et al.\ 2005, \ssr, 120, 95.

\bibitem[Rossi et al.(2019)]{Rossi2019} Rossi, A., Stratta, G., Maiorano, E., et al.\ 2019, arXiv e-prints, arXiv:1901.05792.

\bibitem[Rosswog et al.(2003)]{rosswog03} Rosswog, S., Ramirez-Ruiz, E., \& Davies, M.~B.\ 2003, \mnras, 345, 1077 

\bibitem[Rosswog(2005)]{rosswog05} Rosswog, S.\ 2005, \apj, 634, 1202.

\bibitem[Rowlinson et al.(2010)]{gcn100216a} Rowlinson, A., Page, K., \& Lyons, N.\ 2010, GRB Coordinates Network, Circular Service, No.~10435, \#1 (2010), 10435, 1

\bibitem[Ruffert \& Janka(1999)]{rj99} Ruffert, M., \& Janka, H.-T.\ 1999, \aap, 344, 573 

\bibitem[Ryan et al.(2019)]{Ryan2019} Ryan, G., van Eerten, H., Piro, L., \& Troja, E.\ 2019, arXiv e-prints, arXiv:1909.11691.

\bibitem[\protect\citeauthoryear{Sari, et al.}{1998}]{Sari1998} Sari R., Piran T., Narayan R., 1998, ApJ, 497, L17

\bibitem[\protect\citeauthoryear{Sari \& Esin}{2001}]{Sari2001} Sari R., Esin A.~A., 2001, ApJ, 548, 787

\bibitem[Shibata \& Taniguchi(2011)]{shibata11} Shibata, M., \& Taniguchi, K.\ 2011, Living Reviews in Relativity, 14, 6.

\bibitem[Siegel \& Ciolfi(2016)]{Siegel2016} Siegel, D.~M., \& Ciolfi, R.\ 2016, \apj, 819, 15.

\bibitem[Sylos Labini(2011)]{sylosL2011} Sylos Labini, F.\ 2011, Classical and Quantum Gravity, 28, 164003.

\bibitem[Tanaka et al.(2014)]{tanaka14} Tanaka, M., Hotokezaka, K., Kyutoku, K., et al.\ 2014, \apj, 780, 31 

\bibitem[Tanvir et al.(2005)]{tanvir05} Tanvir, N.~R., Chapman, R., Levan, A.~J., \& Priddey, R.~S.\ 2005, \nat, 438, 991 

\bibitem[The LIGO Scientific Collaboration et al.(2020)]{GW190425} The LIGO Scientific Collaboration, the Virgo Collaboration, Abbott, B.~P., Abbott, R., et al.\ 2020, arXiv e-prints, arXiv:2001.01761.

\bibitem[Thoene et al.(2007)]{thoene07} Thoene, C.~C., Bloom, J.~S., Butler, N.~R., \& Nugent, P.\ 2007, GRB Coordinates Network, 6756, 1

\bibitem[\protect\citeauthoryear{Troja, et al.}{2008}]{gcn080121} Troja, E., Cummings, J.~R., Palmer, D.~M., et al.\ 2008, GCN Report, 118, 1.

\bibitem[\protect\citeauthoryear{Troja, et al.}{2008}]{Troja2008} Troja, E., King, A.~R., O'Brien, P.~T., Lyons, N., \& Cusumano, G.\ 2008, \mnras, 385, L10.

\bibitem[\protect\citeauthoryear{Troja, et al.}{2010}]{Troja10} Troja, E., Rosswog, S., \& Gehrels, N.\ 2010, \apj, 723, 1711.


\bibitem[Troja et al.(2016)]{Troja16} Troja, E., Sakamoto, T., Cenko, S.~B., et al.\ 2016, The Astrophysical Journal, 827, 102.

\bibitem[Troja et al.(2017)]{troja2017} Troja, E., Piro, L., van Eerten, H., et al.\ 2017, \nat, 551, 71 

\bibitem[Troja et al.(2018a)]{Troja2018a} Troja, E., Ryan, G., Piro, L., et al.\ 2018, Nature Communications, 9, 4089.

\bibitem[Troja et al.(2018b)]{Troja2018b} Troja, E., Piro, L., Ryan, G., et al.\ 2018, \mnras, 478, L18.

\bibitem[Troja et al.(2019)]{Troja2019a} Troja, E., van Eerten, H., Ryan, G., et al.\ 2019, \mnras, 489, 1919.

\bibitem[\protect\citeauthoryear{Troja, et al.}{2019}]{Troja2019b} Troja, E., Castro-Tirado, A.~J., Becerra Gonz{\'a}lez, J., et al.\ 2019, \mnras, 489, 2104.

\bibitem[\protect\citeauthoryear{Tunnicliffe, et al.}{2014}]{Tunnicliffe2014} Tunnicliffe R.~L., et al., 2014, \mnras, 437, 1495

\bibitem[Wanderman \& Piran(2015)]{wp15} Wanderman, D., \& Piran, T.\ 2015, \mnras, 448, 3026 

\bibitem[Wen et al.(2013)]{wen2013} Wen, X.-Q., Wu, H., Zhu, Y.-N., et al.\ 2013, \mnras, 433, 2946.

\bibitem[Wollaeger et al.(2018)]{Wollaeger2018} Wollaeger, R.~T., Korobkin, O., Fontes, C.~J., et al.\ 2018, \mnras, 478, 3298.

\bibitem[Xie \& MacFadyen(2018)]{Xie2018} Xie, X., Zrake, J., \& MacFadyen, A.\ 2018, \apj, 863, 58.

\bibitem[Yue et al.(2018)]{yue18} Yue, C., Hu, Q., Zhang, F.-W., et al.\ 2018, \apjl, 853, L10 

\bibitem[Zhang et al.(2018)]{zhang18} Zhang, B.-B., Zhang, B., Sun, H., et al.\ 2018, Nature Communications, 9, 447 

\bibitem[\protect\citeauthoryear{Zou et al.}{2009}]{Zou2009} Zou Y.-C., Fan Y.-Z., Piran T., 2009, MNRAS, 396, 1163



\end{thebibliography}
\end{document}